\documentclass[useAMS,usenatbib]{mn2e}

\usepackage{graphicx}
\usepackage{txfonts}
\usepackage{natbib}
\title[Variable Stars in the Cetus dSph]{Variable Stars in the Cetus dSph Galaxy:
          Population Gradients and Connections with the Star Formation History
\thanks{Based on observations made with ESO Telescopes at the Paranal
 Observatories under programme ID 081.D-0775.}}
\author[M. Monelli et al.]
{M. Monelli$^{1,2}$\thanks{E-mail: monelli@iac.es},
E.~J. Bernard$^{3}$,
C. Gallart$^{1,2}$,
G. Fiorentino$^ {4}$,
I. Drozdovsky$^{1,2}$,
\newauthor
A. Aparicio$^{1,2}$,
G. Bono$^{5,6}$,
S. Cassisi$^{7}$,
E.~D. Skillman$^{8}$,
P.~B. Stetson$^{9}$ \\
$^{1}$Instituto de Astrof\'{i}sica de Canarias, Calle V\'ia L\'actea s/n, E-38205 La Laguna, Tenerife, Spain \\
$^{2}$Departamento de Astrof\'{i}sica, Universidad de La Laguna, E-38200 Tenerife, Spain \\
$^{3}$SUPA, Institute for Astronomy, University of Edinburgh, Royal
      Observatory, Blackford Hill, Edinburgh EH9 3HJ, UK \\
$^{4}$INAF-Osservatorio Astronomico di Bologna, via Ranzani 1, 40127 Bologna, Italy \\
$^{5}$Universit{\'a} di Roma Tor Vergata, Via della Ricerca Scientifica 1, 00133 Roma, Italy \\
$^{6}$INAF-Osservatorio Astronomico di Roma, Via Frascati 33, 00040 Monteporzio Catone, Roma, Italy \\
$^{7}$INAF-Osservatorio Astronomico di Teramo, via M. Maggini, 64100 Teramo, Italy \\
$^{8}$Minnesota Institute for Astrophysics, University of Minnesota, Minneapolis, MN 55455, USA \\
$^{9}$NRC Herzberg Institute for Astrophysics, 5071 West Saanich Road, Victoria, BC V9E 2E7, Canada \\
}

\begin{document}

\date{Accepted 2012 January 11; in original form 2011 November 27}

\pagerange{\pageref{firstpage}--\pageref{lastpage}} \pubyear{2010}

\maketitle

\label{firstpage}

\begin{abstract}

We investigate the variable star content of the isolated, Local Group, dwarf spheroidal 
galaxy (dSph) Cetus. Multi-epoch, wide-field images collected with the VLT/VIMOS camera
allowed us to detect 638 variable stars (630 RR~Lyrae stars and 8 Anomalous Cepheids),
475 of which are new detections. We present a full catalogue of periods, amplitudes,
and mean magnitudes. Motivated by the recent discovery that the pulsational properties
of the RR~Lyrae stars in the Tucana dSph revealed the presence of a metallicity
gradient within the oldest ($\ga$10~Gyr old) stellar populations, we investigated the
possibility of an analogous effect in Cetus. We found that, despite the obvious radial gradient
in the Horizontal Branch (HB) and Red Giant Branch (RGB) morphologies, both becoming
bluer on average for increasing distance from the center of Cetus, the properties of
the RR~Lyrae stars are homogeneous within the investigated area (out to r$\sim$15$\arcmin$),
with no significant evidence of a radial gradient. We discuss this in connection with the 
star formation history (SFH) previously derived for the two galaxies.  The observed 
differences between these two systems show that even systems this small show a variety 
of early evolutionary histories. These differences could be due to different merger or 
accretion histories.

\end{abstract}


   \defcitealias{bernard09a}{Paper~I}
   \defcitealias{monelli10b}{Paper~II}


\section{Introduction}


\begin{table}
 \centering
 \begin{minipage}{180mm}
  \caption{Summary of Observations\label{tab:tab1}}
  \begin{tabular}{@{}ccccccc@{}}
  \hline
 Date       &  UT Start &Filter& Exposure Time & Seeing      \\
            &           &      &   (s)   & ($\arcsec$) \\
 \hline
 2008-07-28 &  08:54:08 & $B$  &   500   & 0.50 \\
 2008-07-28 &  09:03:18 & $B$  &   500   & 0.50 \\
 2008-07-28 &  09:12:29 & $B$  &   500   & 0.50 \\
 2008-07-28 &  09:23:51 & $V$  &   500   & 0.45 \\
 2008-07-28 &  09:37:30 & $B$  &   500   & 0.50 \\
 2008-07-28 &  09:46:40 & $B$  &   500   & 0.50 \\
 2008-07-28 &  09:55:51 & $B$  &   500   & 0.55 \\
 2008-07-28 &  10:07:13 & $V$  &   500   & 0.55 \\
 2008-08-04 &  06:40:03 & $B$  &   500   & 0.55 \\
 2008-08-04 &  06:49:15 & $B$  &   500   & 0.55 \\
 2008-08-04 &  06:58:25 & $B$  &   500   & 0.55 \\
 2008-08-04 &  07:10:18 & $V$  &   500   & 0.55 \\
 2008-08-04 &  07:21:45 & $B$  &   500   & 0.55 \\
 2008-08-04 &  07:30:56 & $B$  &   500   & 0.50 \\
 2008-08-04 &  07:40:06 & $B$  &   500   & 0.55 \\
 2008-08-04 &  07:51:28 & $V$  &   500   & 0.50 \\
 2008-08-04 &  08:03:08 & $B$  &   500   & 0.55 \\
 2008-08-04 &  08:12:18 & $B$  &   500   & 0.55 \\
 2008-08-04 &  08:21:30 & $B$  &   500   & 0.55 \\
 2008-08-04 &  08:32:54 & $V$  &   500   & 0.50 \\
 2008-08-05 &  06:57:57 & $B$  &   500   & 0.60 \\
 2008-08-05 &  07:07:07 & $B$  &   500   & 0.60 \\
 2008-08-05 &  07:16:18 & $B$  &   500   & 0.65 \\
 2008-08-05 &  07:27:39 & $V$  &   500   & 0.55 \\
 2008-08-05 &  07:37:49 & $V$  &   500   & 0.55 \\
 2008-08-05 &  07:50:53 & $B$  &   500   & 0.60 \\
 2008-08-05 &  08:00:04 & $B$  &   500   & 0.60 \\
 2008-08-05 &  08:09:14 & $B$  &   500   & 0.60 \\
 2008-08-05 &  08:20:38 & $V$  &   500   & 0.65 \\
 2008-08-05 &  08:33:40 & $B$  &   500   & 0.60 \\
 2008-08-05 &  08:42:50 & $B$  &   500   & 0.65 \\
 2008-08-05 &  08:52:01 & $B$  &   500   & 0.65 \\
 2008-08-05 &  09:03:22 & $V$  &   500   & 0.55 \\
 2008-08-05 &  09:16:36 & $B$  &   500   & 0.55 \\
 2008-08-05 &  09:25:47 & $B$  &   500   & 0.60 \\
 2008-08-05 &  09:34:58 & $B$  &   500   & 0.65 \\
 2008-08-05 &  09:46:20 & $V$  &   500   & 0.60 \\
 2008-08-06 &  07:14:09 & $B$  &   500   & 0.55 \\
 2008-08-06 &  07:23:20 & $B$  &   500   & 0.50 \\
 2008-08-06 &  07:32:30 & $B$  &   500   & 0.50 \\
 2008-08-06 &  07:44:04 & $V$  &   500   & 0.50 \\
 2008-08-06 &  07:56:44 & $B$  &   500   & 0.50 \\
 2008-08-06 &  08:05:55 & $B$  &   500   & 0.45 \\
 2008-08-06 &  08:15:06 & $B$  &   500   & 0.50 \\
 2008-08-06 &  08:26:30 & $V$  &   500   & 0.45 \\
 2008-08-06 &  08:38:11 & $B$  &   500   & 0.55 \\
 2008-08-06 &  08:47:22 & $B$  &   500   & 0.55 \\
 2008-08-06 &  08:59:30 & $B$  &   500   & 0.55 \\
 2008-08-06 &  09:08:41 & $B$  &   500   & 0.55 \\
 2008-08-06 &  09:17:51 & $B$  &   500   & 0.55 \\
 2008-08-06 &  09:29:15 & $V$  &   500   & 0.55 \\
 2008-08-07 &  08:37:59 & $B$  &   500   & 0.50 \\
 2008-08-07 &  08:47:09 & $B$  &   500   & 0.50 \\
 2008-08-07 &  08:56:20 & $B$  &   500   & 0.55 \\
 2008-08-07 &  09:07:41 & $V$  &   500   & 0.55 \\
 2008-08-07 &  09:20:18 & $B$  &   500   & 0.60 \\
 2008-08-07 &  09:29:28 & $B$  &   500   & 0.65 \\
 2008-08-07 &  09:38:39 & $B$  &   500   & 0.65 \\
 2008-08-07 &  10:01:25 & $V$  &   500   & 0.65 \\
 2008-08-08 &  05:19:34 & $B$  &   500   & 0.80 \\
 2008-08-08 &  05:28:44 & $B$  &   500   & 0.80 \\
 2008-08-08 &  05:37:54 & $B$  &   500   & 0.75 \\
 2008-08-08 &  05:49:21 & $V$  &   500   & 0.80 \\
\hline
\end{tabular}
\end{minipage}
\end{table}


\begin{figure}
 \includegraphics[width=8cm]{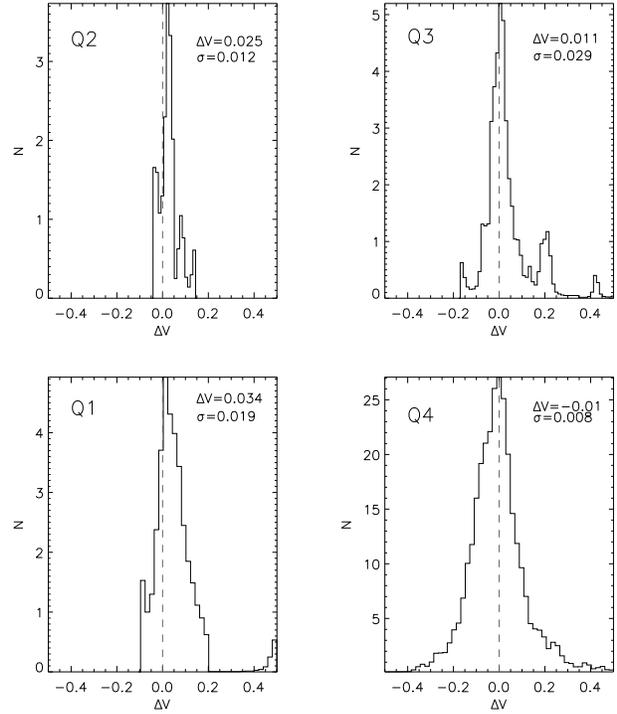}
 \caption{Magnitude residuals in the $V$-band for the stars in common between our
 photometry and that of \citet{mcconnachie06}.}
 \label{fig:calib}
\end{figure}

The galaxies of the Local Group (LG) provide the opportunity to study galaxy evolution
through the analysis of the resolved stellar populations.
These can be investigated in great detail using complementary studies of:
{\em i)} deep color-magnitude diagrams (CMD), with high fidelity photometry
reaching the oldest main sequence turn-off (MSTO), which are crucial to
derive the lifetime star formation history (SFH) of a stellar system \citep{holtzman99,
dolphin02,gallart05, noel08,iacpop, noel09,cignoni10, hidalgo11}; {\em ii)} the properties
of variable stars, which are intrinsically connected to the properties of their corresponding 
stellar population \citep[e.g.,][]{gallart04b}. For example, RR~Lyrae stars are 
unambiguous
tracers of old populations ($> 10$ Gyr), and can provide independent constraints
on the age and metallicity of the host \citep[e.g.,][]{bernard08}; {\em iii)}
the spatial distributions of stellar populations, and the variation of properties
with galactocentric distance, which are useful to constrain the mechanisms affecting dwarf
galaxies at very early epochs \citep{stinson09}; {\em iv)} the chemical and kinematic
properties provided by spectroscopy of individual stars, which can also reveal the presence
of otherwise hidden stellar populations \citep[e.g.,][]{tolstoy04}. This wealth of information
can be combined to derive a comprehensive picture of the evolution of stellar systems.

As part of the Hubble Space Telescope (HST) cycle 14 LCID
project\footnotemark[10]\footnotetext[10]{{\it Local Cosmology from Isolated Dwarfs},
http://www.iac.es/project/LCID}, we obtained
very deep CMDs (V $\sim$ 29~mag) of a sample of five isolated Local Group dwarfs
representative of the dSph (Tucana and Cetus), dIrr (Leo~A and IC~1613) and
transition dIrr/dSph types (LGS~3). These observations allowed us to
derive their lifetime SFHs (\citealt{cole07}, \citealt{monelli10b}, hereafter
Paper II, \citealt{monelli10c, hidalgo11}), and to characterize the properties
of their short period variables (Cetus and Tucana: \citealt{bernard08, bernard09a},
hereafter Paper I; LGS~3 and Leo~A: \citealt{bernard09b}; IC~1613: \citealt{bernard10}).

In particular, Cetus and Tucana present an interesting pair of dwarf galaxies
characterized 
by old ($>$9~Gyr) and metal-poor (Z$<$0.001) stars. Located at more than 700~kpc from
both the Milky Way and M31, they are the two most isolated dSph galaxies in the LG.
Despite their location, they do not host any intermediate-age or young population
\citep{monelli10c}, and therefore they are outliers in the density-morphology
relation \citep{vandenbergh99}. Based on measurements of their radial velocities
\citep{lewis07,fraternali09}, it is believed that they spent most of their life in
isolation, and that they experienced at most one interaction with giant galaxies
in the LG. For these reasons, Cetus and Tucana provide excellent laboratories to study the
mechanisms affecting the early evolution of dSph systems.

\citet{bernard08} presented the intriguing case of Tucana,
where strong gradients {\it within} the old population were observed
in the properties of both the variable and the non-variable stars. Specifically, 
it was observed that:

{\it a)} The color spread of Tucana's Red Giant Branch (RGB), especially for the brightest 
stars, is larger in the inner region than in the outskirts. This is due to the lack 
of red RGB stars in the outer region, where the RGB is predominantly populated at its 
blue edge;

{\it b)} the red part of Tucana's HB is present in the inner region but is sparsely
populated in the external one, while the blue part is present everywhere
across the galaxy;

{\it c)} Tucana's MSTO is thicker in the inner region than in
the outer one, suggesting a larger age spread of the populations in the center
of Tucana.

This empirical evidence indicates the presence of both age and metallicity
gradients in Tucana. The MSTO region indicates that the oldest stars 
are present over the whole body of the galaxy, while the younger stars are more 
centrally concentrated, as commonly observed in Local Group galaxies, e.g., in 
Sculptor \citep{tolstoy04} and Carina \citep{io}. Additionally, the RGB and 
HB morphologies support the idea that the older stars are on average 
relatively metal-poor
(blue RGB, bright blue HB), while the slightly younger stars show enhanced
metallicities (red RGB, faint red HB).

These gradients also show up very clearly in the properties of the 358 RR~Lyrae stars
discovered in Tucana \citep{bernard08}. Thanks to the surprisingly large spread
in the luminosity of the HB in the region of the instability strip, we studied two
components, a bright one and a faint one, finding that they
have different pulsational properties. The corresponding period-amplitude
diagram discloses interesting features. We identified two distinct sequences
corresponding to the fundamental mode RR~Lyrae (RR$ab$), characterized by a
different mean period and a different dispersion around the fit.
In particular, we observed that the stars belonging
to the brighter group have a longer mean period, as predicted by the non-linear 
theory of stellar pulsation
\citep{bono97}. This was also obvious in the mean period of the first-overtone
RR~Lyrae stars (RR$c$);  the ``bell-shaped'' distribution is shifted toward longer
periods for the brighter sample. According to the stellar evolution models, the
luminosity of the HB is a function of the metal content of the stars, with the brighter
stars more metal-poor than the fainter stars \citep{cassisi04}. Note that,
despite the fact that stars evolve from the zero age HB locus to higher  
luminosities \citep{iben70}, this effect cannot account for the observed distribution 
of stars in the CMD of Tucana.
The fainter, more metal-rich RR~Lyrae stars were also found to be also more
centrally concentrated than the brighter ones, thus supporting the idea that a
younger, more metal-rich population is present in the center of Tucana, together
with an older population.

\begin{figure}
 \includegraphics[width=7cm]{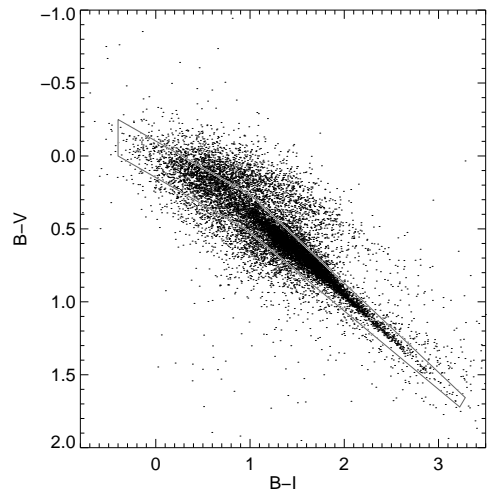}
 \caption{Color-color ($B-V$ vs. $B-I$) diagram. The region enclosed in the box
 was adopted to select the bona fide Cetus stars.}
 \label{fig:colcol}
\end{figure}

Cetus, the other dSph of our sample, appears to host a more homogeneous sample
of RR~Lyrae stars \citepalias{bernard09a}, with a well-defined period-amplitude
relation with a small dispersion. This suggests that the evolutionary history
of Cetus was different from that of Tucana. Indeed, this conclusion is supported
both by the SFHs we derived for the two galaxies \citepalias{monelli10b}, and
by the analysis of the RGB bump \citep{monelli10a}. The SFH of Tucana displays
a stronger initial burst at the oldest possible epochs ($>$ 12.5 Gyr ago), while
Cetus experienced a milder initial episode with the main peak delayed by $\approx$
1 Gyr. Moreover, the RGB of Tucana shows two well defined RGB bumps, associated
with two different populations of slightly different metallicity, in excellent
agreement with the properties of the RR~Lyrae stars. This feature is not present 
in our HST data of Cetus, again supporting a more uniform dominant population in 
this galaxy.

Unfortunately, the ACS field of view allows a very limited radial coverage of Cetus,
$\sim$0.3\% of the area inside the tidal radius.  In contrast, Tucana was 
sampled well beyond its core radius, and partially out to the tidal radius. 
Despite this, \citetalias{bernard09a} showed that 
the morphology of the HB in Cetus changes with radius, with the red stars radial profile
steeper than that of the blue stars. This again indicates that the younger stars 
are more centrally concentrated than the older, more uniformly distributed, stars.

In this paper, we investigate the occurrence of radial gradients in the old
populations of Cetus using the properties of both RR~Lyrae and non-variable stars,
derived from wide-field
observations collected using VIMOS at the VLT. The paper is organized as follows:
\S \ref{sec:obse} summarizes the data acquisition, reduction, and calibration.
In \S \ref{sec:variables} we describe the identification of variable stars and
the derivation of their pulsational parameters. \S \ref{sec:rrl} and \S \ref{sec:cep}
summarize the properties of the RR~Lyrae stars and Anomalous Cepheids detected,
respectively. The radial gradients of both the variable stars and the non-variable 
stars are
presented in \S \ref{sec:radial}. The discussion (\S \ref{sec:discussion}) and
our conclusions (\S \ref{sec:conclusions}) close the paper.


\section[]{Data acquisition and reduction}\label{sec:obse}

 The analysis of variable stars in the Cetus dSph galaxy presented in this paper
 is based on data collected in service mode using the
 VIMOS camera mounted on the VLT (ESO program 081.D-0775, P.I.: E.~J. Bernard). The
 images were acquired during six nights between July 27 and August 08, 2008 (see
 Table~\ref{tab:tab1} for a complete log). The observations were grouped in observing
 blocks (OBs) of 3 images in $B$ and 1 in $V$ of 500 s each. 
 Unfortunately,
 the stringent constraints on seeing to reach the required photometric depth
 prevented the completion of the originally proposed project.  We obtained
 47 $B$ and 16 $V$ images (of the originally proposed 75 $B$ and 25 $V$ images),
 though with an excellent
 average seeing of $\sim$0.6$\arcsec$. Therefore, the total exposure times are
 23,500 and 8,000 sec in the $B$ and $V$ bands, respectively. 

\begin{figure*}
 \includegraphics[width=8cm]{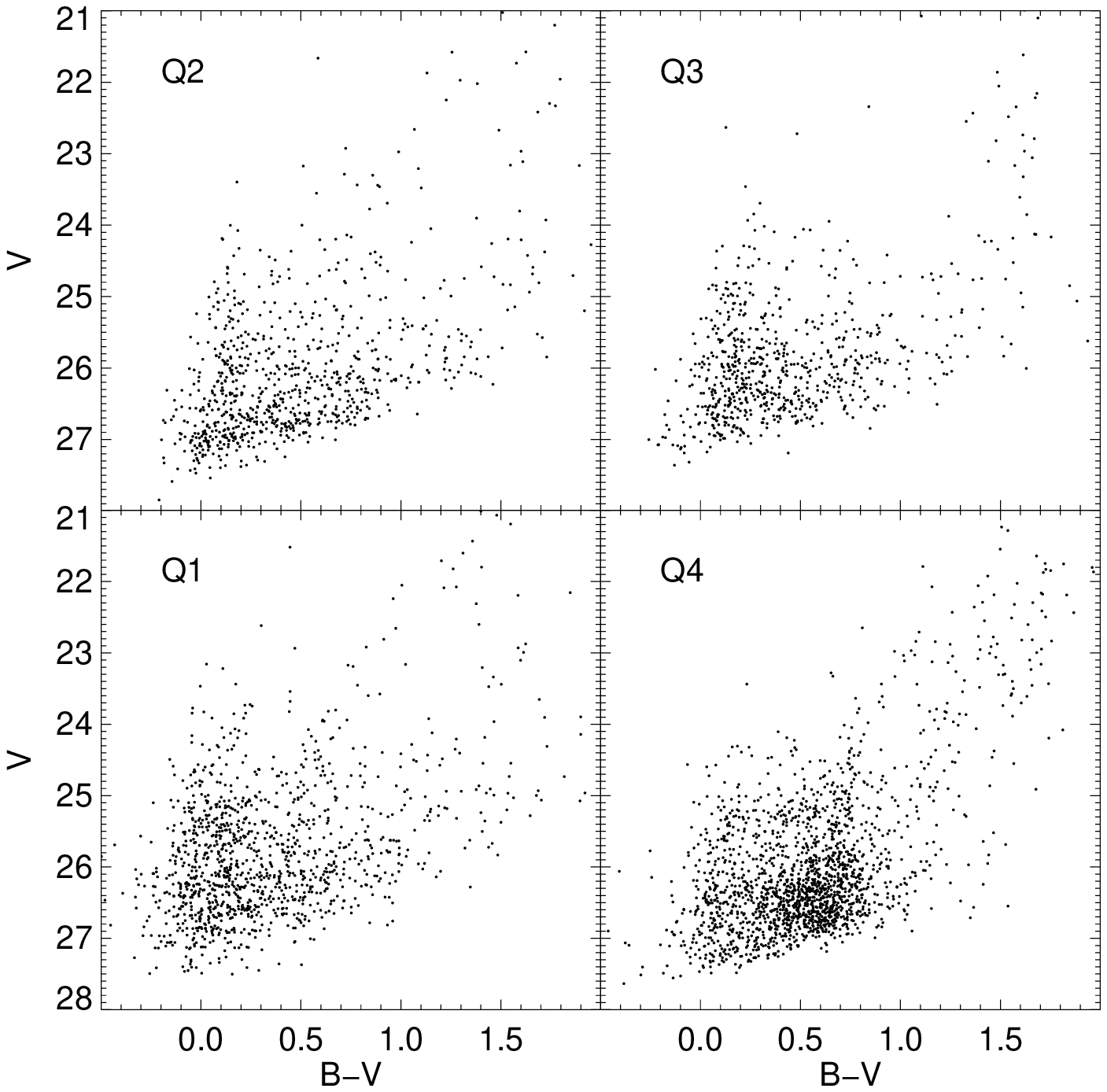}
 \includegraphics[width=8cm]{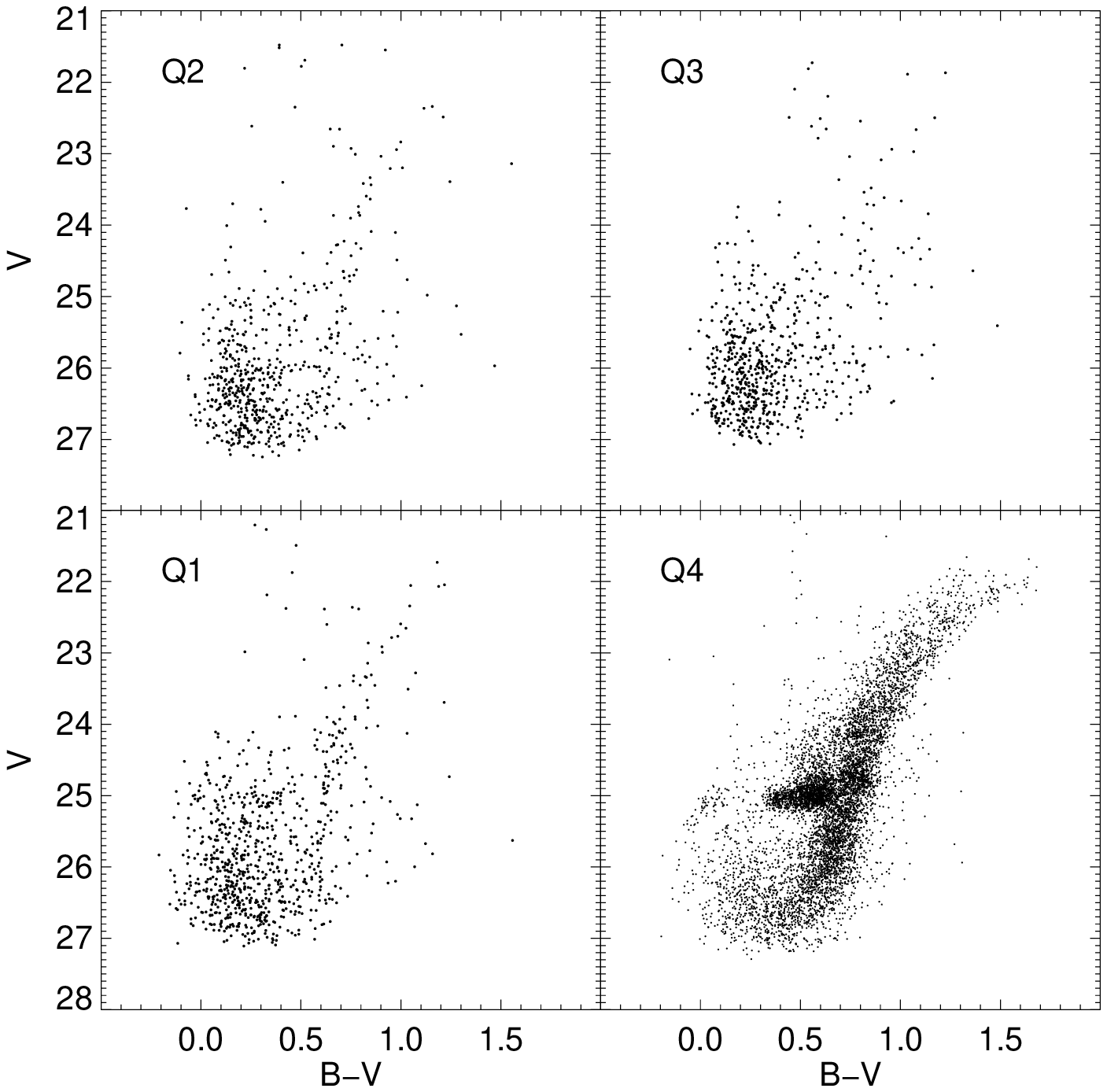}
 \caption{{\itshape Left -} CMD of the objects rejected using the color-color plane
 presented in Figure~\ref{fig:colcol}. The four panels corresponding to the four
 VIMOS chips are presented, as labeled. Note that no obvious features are evident.
 {\itshape Right -} Same as left panels, but the bona fide Cetus stars are presented.
 The RGB and the HB are the main features. Note that Cetus is centered on chip~4.
 \label{fig:clean}}
\end{figure*}

Because of the large gaps between the four VIMOS chips, the galaxy was centered
 in one of the quadrants (chip~4) to cover the center and most of its main body.
 This allowed sampling of the major and minor axes out to $\sim$15$\arcmin$ from
 the center in the North-East direction, corresponding to 3.4~kpc at a distance 
 of 780$\pm$40~kpc \citepalias{bernard09a}. The location of chip~4 also provided 
 almost full coverage of the ACS field-of-view from \citetalias{bernard09a}.
 This allowed us to match the VIMOS and ACS photometry, from which we
 could estimate the completeness of the variable stars detected (see
 section~\ref{sec:compl}), and compare the mean pulsational properties derived
 in the two surveys (see section~\ref{sec:vimosacs}).

We adopted the images processed by the ESO pipeline, and reduced independently each
image of the four chips. The photometry was performed using DAOPHOT IV and
ALLFRAME \citep{alf}. Individual PSFs were modeled using bright stars
covering the whole image, in order to sample the shape variation across the field.
Depending on the total number of stars in each chip, the average number of PSF
stars ranges from $\approx 50$ to $\approx 200$.

 For each chip the input list of stars for ALLFRAME was created using the median
 image. Due to the heavy contamination of background galaxies, especially in the
 three external chips, particular attention has been devoted to clean the input
 list, both through visual inspection of the images and using the shape parameter
 provided by DAOPHOT ($sharpness$). However, the efficiency of the latter procedure 
 is hampered  by the strong variation of the PSF across the field, which causes the 
 stars to be considerably elongated at the corners of  the chips, as well as by 
 crowding in the central chip. For these reasons, the final catalogs still contain a 
 significant number of unresolved galaxies (see below).
 
 In order to further clean the CMD from non-stellar sources we selected the stars based 
 on their position in a color-color plane ($B-V$
 vs. $B-I$), as detailed in \S \ref{sec:cmd}. To do this, we supplemented the 
 present VIMOS data with $I$ images from the Subaru archive. These data consist of 10 $I$ 
 images, each with exposure time of 240 s, collected with the Suprime Cam on Aug 2005. 
 The typical seeing was of the order of 0.6$\arcsec$, therefore homogeneous
 with that of the VIMOS observations. Despite
 the larger area covered by this instrument ($34\arcmin\times27\arcmin$) compared 
 to that  of VIMOS ($18\arcmin\times16\arcmin$), the short time baseline of few hours
 did not allow us to use these data to extend the search for variable stars.
 Therefore, the $I$ data have been used only to build the color-color plane used
 to clean the CMD from foreground and background objects. The image pre-reduction
 was performed using the standard {\itshape SDFRED1} pipeline available for the
 instrument. The PSF modeling and photometry were carried out following the same
 recipe as for the VIMOS data. Note that the four VIMOS chips overlap with
 different Subaru chips. Therefore, for the sake of simplicity, we ran four independent
 ALLFRAME analyses on the overlapping areas only.

    \subsection[]{Photometric Calibration}\label{sec:calibration}

 The photometric calibration was based on standard fields observed as part of
 the routine calibration plan that ESO offers for service mode observations.
 We selected the data collected during the night of August 3, 2008, since
 it was the only photometric night when two fields were observed (PG1633 and
 PG0231) at slightly different airmasses. One image per band per field was
 secured, with exposure time of 2~s. Unfortunately, the difference in airmasses
 (1.16 and 1.30) was too small to obtain a reliable estimate of the atmospheric
 extinction coefficients. For this reason, we adopted the mean coefficients for
 P81 from the VIMOS web page\footnotemark[11].
\footnotetext[11]{http://www.eso.org/observing/dfo/quality/VIMOS/qc/zeropoints.html}

We performed photometry with various apertures to establish the optimal one. The
catalogues were then matched with the lists available on P.B. Stetson's web
page\footnotemark[12]. After correcting for the airmass and exposure time, we
estimated the calibration curves. Due to the low SNR of the VIMOS images, the
 $\Delta${\it B} = {\it B$_{std}$ -- B$_{obs}$} shows a relatively large dispersion, 
 and since it was not possible  to firmly establish any color dependency we estimated
 a simple zero-point to shift to the standard system. On the other hand, the
 $\Delta V$ = {\it V$_{std}$ -- V$_{obs}$} presents a much smaller dispersion,
 and does not show any obvious color trend. Therefore, as for the $B$-band we
 only applied a zero-point. Note that we repeated this procedure using the 
 two standard fields separately, and
then using all the stars together, obtaining the same results within the errors.
\footnotetext[12]{http://www3.cadc-ccda.hia-iha.nrc-cnrc.gc.ca/community/STETSON/standards/}

\begin{figure}
 \includegraphics[width=9cm]{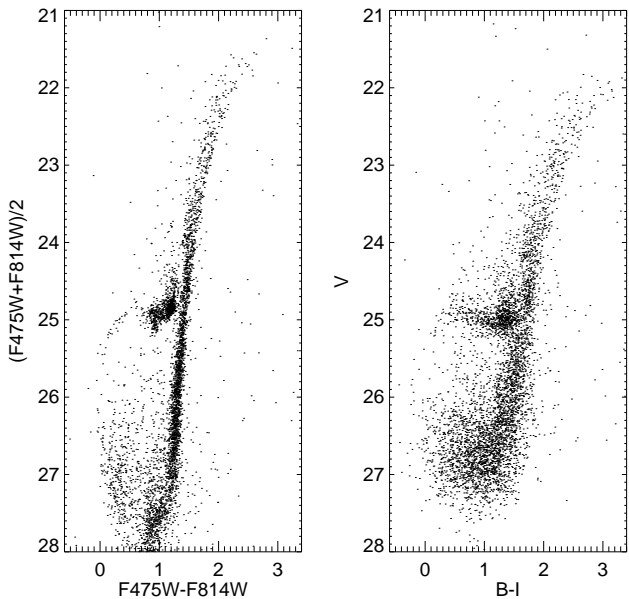}
 \caption{Comparison of the ACS ({\itshape left}) and VIMOS ({\itshape right}) CMDs, 
 showing only the stars in common.
 Note the effect of blending, that pushes very faint stars to be detected in the VIMOS
 photometry.}
 \label{fig:vimos_acs}
\end{figure}

To apply this calibration to the individual catalogues of all the Cetus images,
we applied the following steps. First, we picked two reference images, one for
the {\it B} and one for the {\it V} band, both observed during the same night
as the standard fields. We then rescaled each individual catalogue to the
reference one. This was done applying a shift in magnitude, estimated using
the brightest stars in common with the reference list. Once every catalogue
had been homogenized to the photometric system of the reference images, we
applied the calibration previously described to each of them.

\begin{figure}
 \includegraphics[width=8cm]{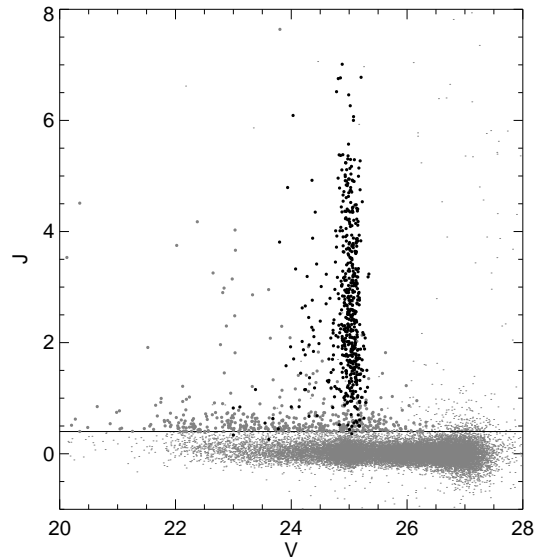}
 \caption{Variability index as a function of magnitude for chip~4. The variable
 stars, discarded candidate variables, and non-variable stars are shown
 as filled black circles, filled gray circles, and dots, respectively. The
 horizontal line shows the cut used to extract the candidates. See text for
 details.}
 \label{fig:varindex}
\end{figure}

 Figure~\ref{fig:calib} shows the residuals for the stars in common between our
 photometry in the $V$-band and that of \citet{mcconnachie06}. There seems to be
 small offsets, which are different for each chip and most likely due to the
 poor quality of our standard field images. However, none of the results
 presented in this paper rely on high precision photometry. We thus consider the
 quality of this calibration to be sufficient for our purposes.

    \subsection[]{Color-Magnitude Diagram}\label{sec:cmd}

Figure~\ref{fig:colcol} presents the ($B-V$ vs. $B-I$) color-color diagram. We applied 
a cut in the sharpness parameter ($|sharp| <$ 0.3) and photometric error ($\sigma_{V}$, 
$\sigma_{B}$, $\sigma_{I} < $0.3 mag). Two
sequences appear in this plane. The first is tight and well defined, extending from 
$B-I\sim1$ to $B-I\sim3.5$. This is composed of Cetus RGB stars. The second one, 
less concentrated, is characterized by smaller $B-V$ for fixed $B-I$, compared to the
RGB stars. In this region, background galaxies are expected \citep[e.g.,][]{roccavolmerange08}.
The plot also shows the area used to isolate the bona fide Cetus stars. The limits at 
the blue side have been fixed inspecting both the ($V$, $B-V$) and ($V$, $B-I$) CMD of 
chip~4, where the significant number of blue stars helped to define the locus occupied 
by the HB on the color-color plane. 
As a cross-check of our approach, we selected a sample of $\sim 20$ candidate blue HB 
stars and visually inspected them on both the VIMOS and Subaru images, concluding that
95\% of them present a stellar profile, while the nature of only one object was unclear.
Moreover, we verified that this selection did not affect the sample of variable stars:
in fact, out the 638 detected variables, only one was lost probably due to its proximity 
to the Subaru chip edge.

Figure~\ref{fig:clean} shows the $V$, $B-V$ CMD of the rejected (left) and selected (right)
objects. In both cases, the CMD of the four chips is shown.
The left panels show that very few Cetus RGB stars have been removed from chip~4, while
no structures typical of the CMD of a stellar system are evident in the CMDs of 
chips~1,2,3. The right panels show that our photometry reaches $\sim$1.5 mag 
 below the HB, down to $V \sim 27$ mag, which  is comparable in depth to the WFPC2 
 parallel observations presented in \citetalias{bernard09a}. The CMD of chip~4, 
 centered on Cetus, shows that the RGB and the HB are the most evident features. 
 The other panels show clear evidence of the presence of both the RGB and HB in 
 chip~1 and chip~2, while these are only barely visible in chip 3. A residual component 
 of blue, faint objects ($V > 25, B-V \le 0.3$) is still present, and particularly 
 evident in the outer chips. It is certainly possible that a number of background 
 galaxies are still polluting the CMD, but other sources might be present.

First, we investigated the level of contamination from foreground Galactic stars,
performing a few tests using both the TRILEGAL code\footnotemark[13]
\citep{girardi05}\footnotetext[13]{http://stev.oapd.inaf.it/cgi-bin/trilegal}
and the Pisa Galactic Model code \citep{castellani02}. We verified that both codes 
predict a very small number of foreground objects (Galactic
blue HB or white dwarf stars) in this region of the CMD ($<$10 in the whole
VIMOS field).  This is not surprising, since the high Galactic latitude of Cetus 
($b = -72.9 \degr$) implies that a very small portion of Galactic disk is observed along 
the line of sight. 

\begin{figure}
 \includegraphics[width=8cm]{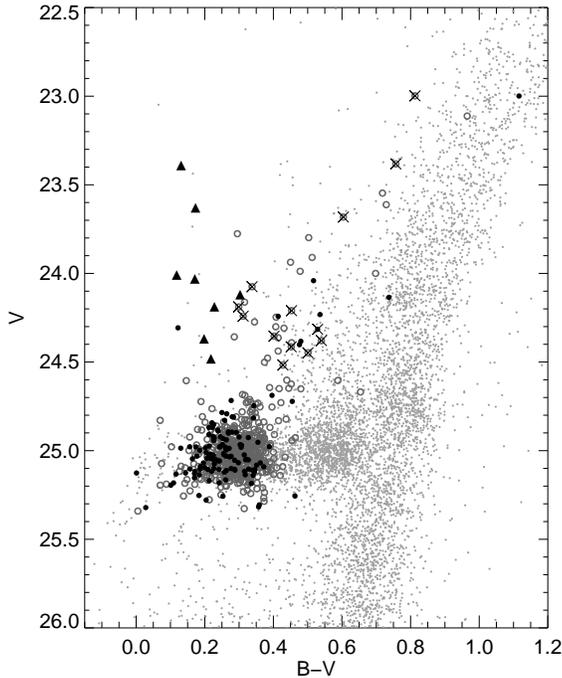}
 \caption{CMD of chip~4, where the variable stars of all four chips have been
 over-plotted: RR$ab$, RR$c$, and Anomalous Cepheids are shown as red open circles, 
 black filled circles,
 and filled triangles, respectively. The crossed circles indicate bona fide
 RR~Lyrae stars from \citetalias{bernard09a} which are blended in the VIMOS
 observations and therefore appear much brighter.}
 \label{fig:cmd_vars}
\end{figure}

Alternatively, they could be stars belonging to Cetus. Based on the SFH results
\citepalias{monelli10b} we can eliminate the possibility that Cetus hosts a 
significant population of intermediate-age 
to young main sequence stars ($\le$5 Gyr old). On the other hand, the analysis of the 
 ACS CMD revealed the presence of a sizable population of Blue Stragglers 
 \citep[BSS;][]{monelli12}, estimated to be on the order of 5,000 stars in the whole galaxy
 body. These stars have colors and magnitudes similar to those of a 3 to 5~Gyr old 
 population. However, it is important to note that the luminosity of the brightest 
 BSSs is significantly fainter than the typical blue HB star. This means that, even
 if BSSs are certainly present (but are expected to be very few in the outer chips)
 they do not affect the HB nor the RGB morphology.

Finally, crowding is also expected to artificially shift objects outside the main
Cetus sequences.
A direct comparison of the VIMOS and ACS photometry is shown in Figure \ref{fig:vimos_acs},
where the CMD of the $\sim 5,300$ stars in common are presented. It is interesting to note
how a number of very faint stars ($V \sim 27$ or fainter) are detected in the VIMOS field
with significantly brighter magnitude. This is certainly due to the effect of crowding, 
which causes the blending of sources. This shows that at least some of the blue objects with 
26 $< V <$27 are the results of such a mechanism, even though the crowding in the outer
regions is expected to be less important.

Overall, the selection performed in the color-color plane removed from the CMD an important
fraction of contaminating field sources, at least at the magnitude level of the HB or brighter. 
The residual sources, whatever their nature, have negligible
effect on the HB and RGB morphology. Therefore, we assume that the analysis presented in the following
sections (see e.g. \S \ref{sec:radial}) is based on a bona fide sample of Cetus stars.
However, we stress that the search for variable stars, detailed in the next section, was 
performed on the complete, unselected $B$,$V$ VIMOS catalogue.


\section[]{Variable stars}\label{sec:variables}

    \subsection[]{Identification and period search}\label{sec:period}

 The search for variable stars is based on the Stetson variability index $J$
 \citep{stetson96}.
 Figure~\ref{fig:varindex} shows the distribution of $J$ as a function of V-band
 magnitude for chip~4. Stars with at least 10 measurements in each band and
 $J>$~0.4 were considered candidate variables, and are shown as larger gray and
 black symbols. This value of the index was chosen in order to select most
 variables while keeping the number of false detections low. The location of the
 bona fide variable stars (see below), shown as black symbols, indicate that
 very few have $J\la$~0.8.
 The three variables with $J<$~0.4 have very noisy light curves due to crowding
 but are known to be actual variables from the survey of \citetalias{bernard09a}.
 We discuss the completeness of our sample in section~\ref{sec:compl}.

 This process yielded 413, 538, 160, and 1020 candidates in the four chips.
 Unfortunately, these numbers include a large fraction of background galaxies, as
 described above, and careful visual inspection of the images and light curves
 was carried-out to reject false detections. Of these, 48, 22, 20, and 548
 turned-out to be actual variable stars for which we could phase the light curve.

 The search for periods was performed with the same algorithm as adopted in
 \citetalias{bernard09a}. It is based on the Fourier analysis \citep{horne86}
 taking into account the information from both bands simultaneously and rejecting
 outliers with large uncertainty. For each variable, we estimated the mean 
 uncertainty and its standard deviation $\sigma$, and rejected individual data
 points with error larger than 3-$\sigma$ above the mean error. This procedure was 
 iterated 5 times. The code allows
 interactive refinement by modifying the period until a tighter light curve is
 obtained. Visual inspection of all of the candidates' light curves was necessary
 because, due to the time sampling, the determination of the period for a number
 of variables was complicated by aliasing problems.

 The classification of the candidates was based on their light-curve morphology
 and position in the CMD. Figure~\ref{fig:cmd_vars} shows the CMD of chip~4 where
 the variables from all four chips have been overlaid. As expected, for the old age
 of this galaxy, most variables are also RR~Lyrae stars, though there is a
 large number of variable stars brighter than V$\sim$24.7. However, visual inspection of
 the stacked image revealed that most of these are RR~Lyrae stars blended with
 RGB stars and therefore appear significantly brighter and redder than the HB. This is
 confirmed by the comparison with the stars in common with the ACS catalog: the
 crossed symbols indicate stars identified as bona fide RR~Lyrae stars in the
 latter, for which the magnitudes were overestimated by up to two magnitudes.
 However, the identification as Anomalous Cepheid or RR~Lyrae star was rather 
 straightforward,  since the former have very smooth light curves with large 
 amplitudes and very small photometric uncertainties, unlike the latter.

\begin{figure*}
 \includegraphics[width=14cm]{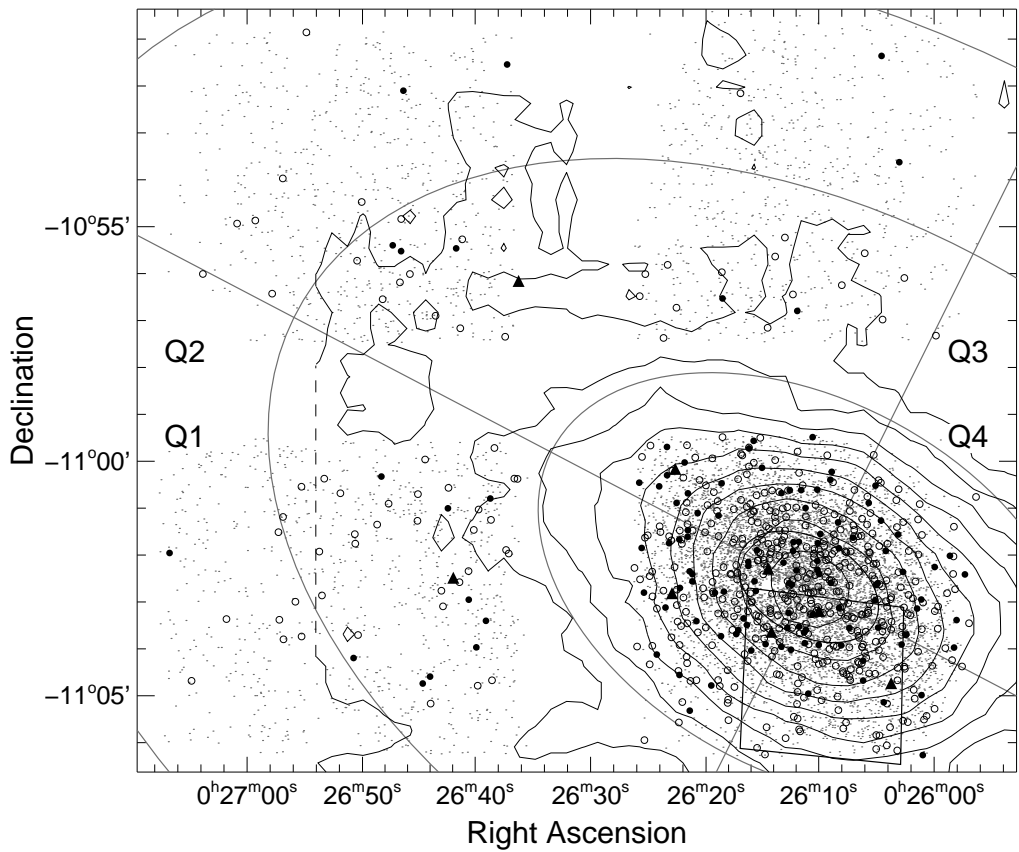}
 \caption{Spatial distribution of stars in the VIMOS field-of-view, where the
  four chips are labeled Q1 to Q4. Fundamental and first-overtone RR~Lyrae stars
  (open and filled circles, respectively) and Anomalous Cepheids (filled
  triangles) are also shown. The black innermost ellipse represents the core
  radius \citep[$r_c=1.3\arcmin \pm 0.1$;][]{mcconnachie06}, while the gray
  ellipses are shown at equivalent radii of 5, 10, and 15$\arcmin$ from the
  center. The isopleth map from \citet{mcconnachie06} is over-plotted.}
 \label{fig:xy}
\end{figure*}

 In total, our catalogue includes 630 RR~Lyrae stars, and 8 Anomalous Cepheids.
 This includes the three RR~Lyrae stars with $J<$~0.4 that were not detected by
 the process described above but recovered after comparison with the sample of
 \citetalias{bernard09a}.
 Roughly 75\% of these are new discoveries (470 out of 630 RR~Lyrae stars, and 5
 out of 8 Anomalous Cepheids). The properties of the RR~Lyrae stars and Anomalous
 Cepheids are detailed in Table~\ref{tab:tab2} (a complete version is available 
 on-line) and 3, respectively. Finding charts for the whole sample of variable 
 stars are presented in the Appendix. Given the relatively low quality of the 
 light curves and, more importantly, the short observational timebase, the 
 periods are given with three significant digits or less.


    \subsection[]{Completeness}\label{sec:compl}

 Since chip~4 almost completely overlaps with the location of the field surveyed
 in \citetalias{bernard09a}, we can estimate the completeness of our sample by
 comparing the variables detected in either catalog in the area in common.
 As expected, all the variables detected in the VIMOS catalog had already been
 discovered in the ACS data, thanks to the much higher resolution and depth. On
 the other hand, 20 RR~Lyrae variables from the latter catalog
 were missing in the present sample. We investigated the reasons for their
 absence and found the following:

\begin{itemize}
\item 11 are missing from the photometric catalog, either because they are
 blended with other stars/galaxies (V003, V004, V012, V028, V075, V078, V165,
 and V169) or located in the corner of chip~4 where the geometric distortions
 produced elongated PSFs (V192, V194, and V195). These were most likely rejected
 from the master list based on their shape parameters or could not be fit by
 ALLFRAME.
\item 5 did not pass the $J>$~0.4 requirement (V002, V006, V059, V106, and V143).
 Their photometry is contaminated by nearby stars/galaxies and therefore have
 very noisy light curves. Using the periods from the ACS survey, we were able to
 fit the light curves and obtain mean magnitudes for the first three, while the
 light curves for the other two are too noisy and are not included in the present
 catalog.
\item 3 have $J>$~0.4 but were initially rejected as non variables because of
 their noisy light curves (V023, V039, V100). Likewise, we could fit the light
 curves of V039 and V100, while V023 is too crowded.
\item V022 lies just outside of the VIMOS field-of-view.
\end{itemize}

\begin{table*}
\centering
\begin{minipage}{130mm}
 \caption{Pulsational Properties of the RR~Lyrae Stars. For those in common with
 the ACS survey, we adopted the same ID. The newly discovered variable stars are
 sorted by increasing R.A. The complete table is available from the on-line edition.
 \label{tab:tab2}}
 \begin{tabular}{@{}ccccccccccc@{}}
 \hline
ID & Type & R.A. & Decl. & Period & log P & $\langle B \rangle$ & A$_{B}$ & $\langle V \rangle$ & A$_{V}$ & Comments\\
  &      & (J2000) & (J2000) & (days) &  &                     &         &                     &         &         \\
\hline
V001 &   ab  & 0:26:02.84  & -11:03:41.1  & 0.667  & -0.176 &  25.397 &  0.535 & 24.943 &  0.488 &           \\
V002 &    d  & 0:26:02.84  & -11:03:53.5  & 0.4    & -0.4   &  24.115 &  0.122 & 22.999 &  0.030 & blended   \\
V005 &   ab  & 0:26:03.22  & -11:06:08.5  & 0.605  & -0.218 &  25.367 &  1.157 & 25.048 &  0.833 &           \\
V006 &   ab  & 0:26:03.19  & -11:04:23.3  & 0.8    & -0.1   &  25.446 &  0.261 & 25.045 &  0.175 &           \\
V007 &   ab  & 0:26:03.20  & -11:04:39.1  & 0.65   & -0.19  &  25.534 &  0.341 & 25.182 &  0.156 & noisy     \\
\hline
\end{tabular}
\end{minipage}
\end{table*}

\begin{table*}
 \centering
 \begin{minipage}{130mm}
  \caption{Pulsational Properties of the Anomalous Cepheids. For the two stars for which
  the classification is based on the morphological approach only, the type is put in
  parenthesis.\label{tab:tab3}}
  \begin{tabular}{@{}ccccccccccc@{}}
  \hline
ID & Type & R.A. & Decl. & Period & log P & $\langle B \rangle$ & A$_{B}$ & $\langle V \rangle$ & A$_{V}$ & Mass \\
   &      & (J2000) & (J2000) & (days) & (days) &              &          &                     &         & M$_{\odot}$ \\
 \hline
V009 & $F$    & 0:26:03.75 & $-$11:04:42.7 & 0.866  & $-$0.062 & 24.423 & 1.221 & 24.120 & 0.945 & 1.1 \\ 
V095 & $FO$   & 0:26:10.06 & $-$11:03:10.7 & 0.647  & $-$0.189 & 23.522 & 0.596 & 23.391 & 0.474 & 1.2 \\ 
V156 & $(FO)$ & 0:26:14.20 & $-$11:03:37.3 & 0.536  & $-$0.271 & 24.567 & 0.611 & 24.369 & 0.503 & n.c.\\ 
V422 & $(F)$  & 0:26:14.49 & $-$11:02:16.9 & 0.57   & $-$0.244 & 24.416 & 0.974 & 24.188 & 0.627 & n.c.\\ 
V564 & $FO$   & 0:26:22.61 & $-$11:00:09.7 & 0.437  & $-$0.360 & 24.127 & 0.847 & 24.009 & 0.639 & 1.0 \\ 
V567 & $FO$   & 0:26:22.89 & $-$11:02:48.1 & 0.472  & $-$0.327 & 24.202 & 1.044 & 24.031 & 0.750 & 1.0 \\ 
V601 & $FO$   & 0:26:36.29 & $-$10:56:10.3 & 0.457  & $-$0.340 & 23.803 & 1.748 & 23.630 & 1.050 & 1.8 \\ 
V623 & $F$    & 0:26:41.99 & $-$11:02:28.4 & 0.426  & $-$0.371 & 24.700 & 1.364 & 24.482 & 1.116 & 1.6 \\ 
\hline
\end{tabular}
\end{minipage}
\end{table*}

\begin{figure}
 \includegraphics[width=9.0cm]{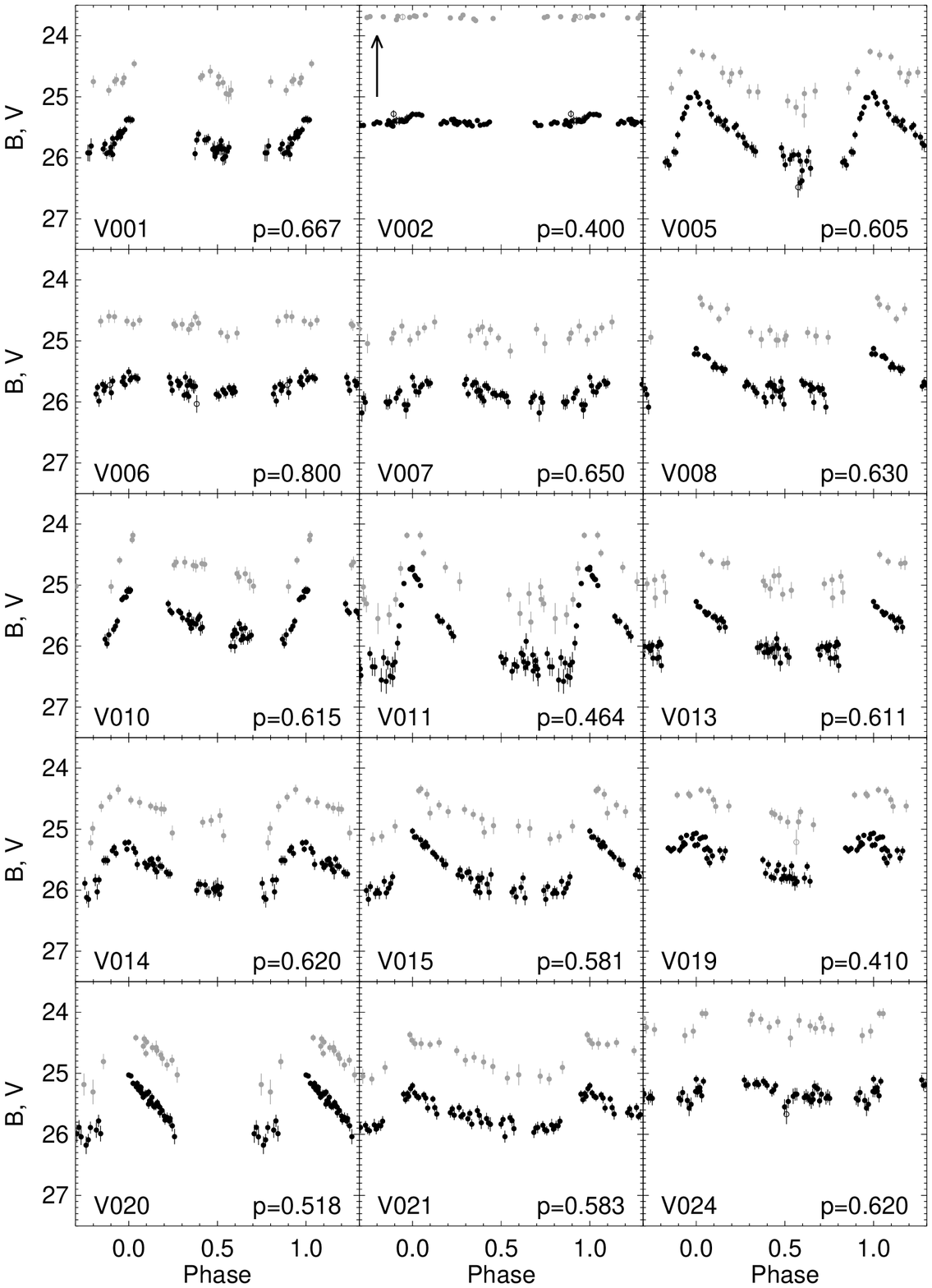}
 \caption{Sample light curves of the RR~Lyrae stars in the $B$ (black) and $V$
 (gray) bands, phased with the period in days given in the lower right corner of
 each panel. Open symbols show the bad data points, i.e., with errors larger
 than 3-$\sigma$ above the mean error of a given star, which were not used in
 the calculation of the period and mean magnitudes. For clarity, the $B$ and $V$
 light curves have been shifted by 0.3 mag down- and upward, respectively. In the
 few cases where the RR~Lyrae stars appear significantly brighter due to blending,
 shown by the arrow in the top left corner of a panel, the light curves have
 been shifted downward by 1 mag.
 All of the light curves are available in the on-line version.}
 \label{fig:rrl_lc}
\end{figure}

 Crowding is thus the main reason limiting the completeness of our variable star
 sample. Assuming the ACS catalog is 100\% complete, this means that the present
 sample is about 89\% complete in the region in common with the former (19
 RR~Lyrae stars missed, out of 172). Since most of the VIMOS field is less
 crowded than this area, this suggests a completeness of $\ga$ 90\% over the
 surveyed field.

 On the other hand, Cetus is a large galaxy and the field-of-view of VIMOS cannot
 cover it completely. Figure~\ref{fig:xy} shows the distribution of stars in
 the field, where open and filled circles mark the position of the RR~Lyrae stars, 
 and filled triangles show the Anomalous Cepheids. The black ellipse shows the core radius,
 while the three gray ones indicate the r=5, 10, and 15$\arcmin$ distance from the center.
 We also over-plotted the isopleth contours from \citet[their Figure~2]{mcconnachie06}. The
 vertical dashed line on the outermost contour marks the eastern edge of their field.
 As shown on Figure~\ref{fig:xy}, not only is VIMOS much smaller than
 the tidal radius \citep[$r_t=32.0\arcmin \pm 6.5$,][]{mcconnachie06}, but it also
 presents wide gaps between the chips. To estimate the total number of RR~Lyrae
 stars expected within the tidal radius of this galaxy, we used their density
 profile assuming that
 they are uniformly distributed in each elliptical bin. Integrating the radial
 profile of RR~Lyrae stars we find an expected total number of $\sim$ 1,200,
 in good agreement with previous estimates based on the much smaller area covered
 by the ACS data \citepalias[$\sim$1,000;][]{bernard09a}.  Thus, we have directly 
 measured the properties of roughly half of all of the RR~Lyrae stars in Cetus.

Concerning the other types of variable stars identified in the ACS survey, 
none of the 4 binary star candidates identified in the ACS photometry were recovered
as variables in this survey. Two of them are non-varying in the VIMOS data (V16, 
V60), probably due to the insufficient sampling of the light curve, while the other 
two (V18 and V146) are too faint to reliably assess their variable nature.
Finally, the long period variable star V17 was not identified as a candidate
variable in this survey, probably due to the lower sensitivity and limited time
baseline.

\begin{figure}
 \includegraphics[width=8cm]{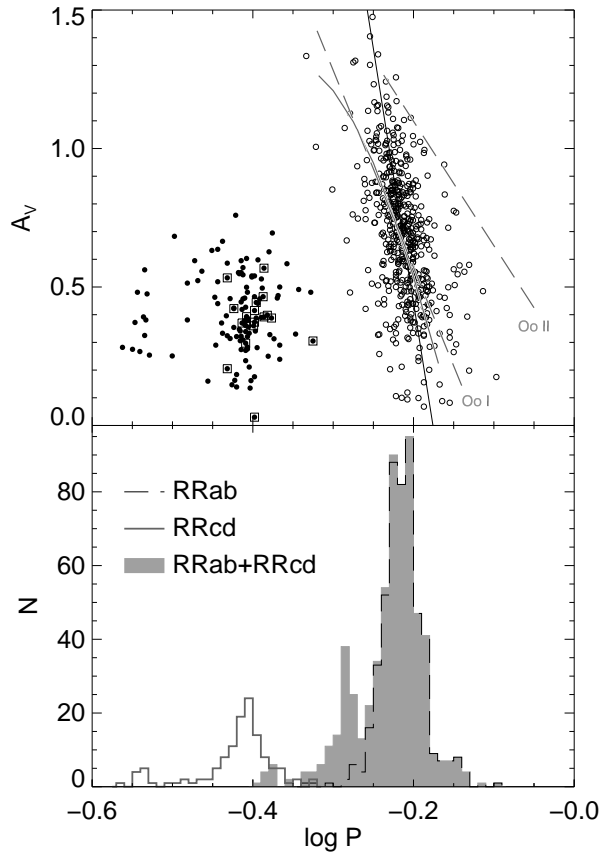}
 \caption{{\it Top:} Period-amplitude diagram for the RR~Lyrae stars in Cetus.
 Open and filled circles represent RR$ab$ and RR$c$, respectively. Square
 symbols show the RR~Lyrae stars identified as RR$d$ in the HST data. The solid
 line is a fit to the period-amplitude of the RR$ab$.
 The solid gray line represents the average distribution of the (unevolved)
 RR$ab$ variables in M\,3, from \citet{cacciari05}, while the dashed gray lines
 delineate the P-A$_{V}$ relations for RR$ab$ stars in OoI and OoII clusters
 \citep{clement00}.
 {\it Bottom:} Period histogram for the RR~Lyrae stars of the top panel. RR$ab$
 and RR$c$ are shown as histograms with dashed and solid lines, respectively, while
 the filled gray histogram represents the fundamentalized RR~Lyrae stars.}
 \label{fig:pa_hist}
\end{figure}

\section[]{RR~Lyrae stars}\label{sec:rrl}

    \subsection[]{Global properties}\label{sec:global}

 From the periods and light curve shapes of the candidates, we identified 506
 RR~Lyrae stars pulsating in the fundamental mode (RR$ab$) and 124 in the
 first-overtone mode or in both modes simultaneously (RR$c$ or RR$d$).
 Unfortunately, the sub-optimal SNR of our individual data points made the
 distinction between the latter two types very challenging. In addition,
 blending tends to increase the noise in the light curves, further complicating
 the discrimination. Therefore, in the following, only the RR$d$ confirmed as
 such in the ACS sample will be considered bona fide double-mode pulsators.
 The light curves of all the RR~Lyrae stars are shown in Figure~\ref{fig:rrl_lc}.

 The top panel of Figure~\ref{fig:pa_hist} presents the period-amplitude diagram
 (P-A$_{V}$) for the RR~Lyrae stars detected in the four chips. The sequence of
 the RR$ab$ is well defined and presents a small dispersion ($\sigma$=0.023
 along the x-axis) around the fit
 (solid black line). However, as already hinted in \citetalias{bernard09a}, we
 find that the slope of the fit is unusually steep (slope=$-$18.3), but similar to 
 the P--A$_{V}$  of Carina shown in \citet{dallora03}. For comparison, the typical
 P--A$_{V}$ relations for RR$ab$ stars in OoI and OoII globular clusters from
 \citet{clement00} are shown as dashed gray lines, and the non-linear fit to the
 unevolved RR$ab$ stars of M\,3 as the solid gray line \citep{cacciari05}. None of
 these provide a good fit to our observed distribution. 

 While the slope does depend on the adopted mixing-length parameter $l/H_p$, in
 the sense that an increase in the efficiency of convection in the star external
 layers leads to a steeper P--A$_{V}$ relation \citep[see][and references
 therein]{bono07}, robust correlations with physical parameters have not been
 established yet and the suggested ones are still controversial
 \citep[e.g.,][]{bono97, casagrande07}.

 We note that a few variable stars with unusually low amplitude ($\sim0.1$mag) 
 appear in the P-A$_{V}$. We verified that they belong to the group of variables
 significantly brighter than the HB (see Figure \ref{fig:cmd_vars}) and are blended in
 the stacked images, and therefore conclude that they are affected by crowding.
 This causes a reduction of the amplitude, shifting the stars downward
 in the P-A$_{V}$ plane. However, we can exclude that this bias is artificially
 steepening the slope of the period-amplitude relation, because the change of
 slope when removing the blended stars is insignificant.
 Moreover, the derived estimate of the slope is perfectly consistent with the
 one obtained from the ACS data \citepalias{bernard09a}, where the crowding has
 a negligible effect.

 Figure~\ref{fig:pa_hist} also shows that the majority of RR$cd$ stars are concentrated 
 around $log P\sim-$0.4~day, although a secondary clump is visible at 
 $log P\sim-$0.54~day. This bi-modality of the RR$cd$ is also obvious in the bottom 
 panel, which shows the histograms corresponding to the stars of the top panel.
 The dashed black and solid gray histograms show the period distributions for RR$ab$ and
 RR$cd$ stars, respectively, while the shaded histogram represents the
 distribution of {\it fundamentalized} periods: the periods of the RR$cd$ stars
 were transformed to their fundamental mode equivalents by adding 0.128 to the
 logarithm of their periods.
 While the bi-modality may partly be due to aliasing for the stars with the
 shorter period, i.e., an alias was chosen instead of the true period, at least
 for one of these we can be sure of the period since it also appears in the ACS
 catalog and cannot be phased with a longer period. 
 It is worth mentioning that there have been claims in the literature about the
 discovery of RR~Lyrae stars possibly pulsating on the second overtone, or RR$e$, which 
 are expected to have a period close to 0.3~day \citep[see e.g.][]{soszynski09,
 soszynski10}, thus similar to the second peak we detected.
 Given the lack of theoretical support and the possible problems with aliasing,
 we cannot safely classify them as RR$e$, so in the present work we assume that
 these stars are bona fide RR$c$ stars. We show sample light curves in
 section~\ref{sec:note}.

\begin{figure}
 \includegraphics[width=8cm]{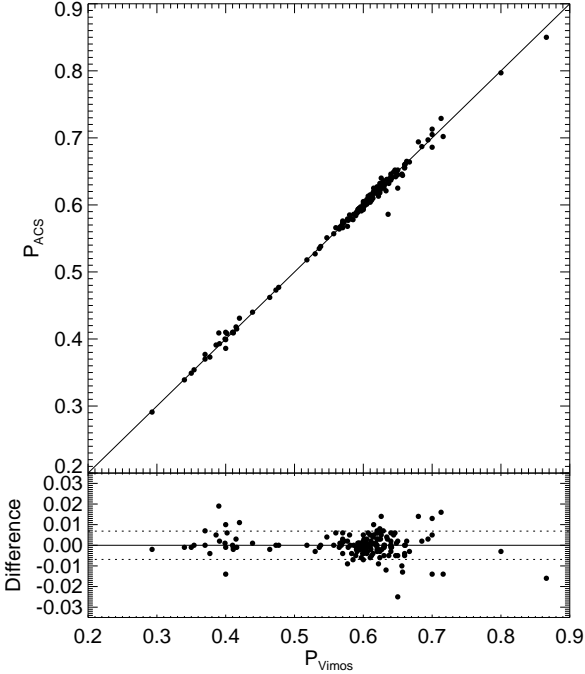}
 \caption{Period difference for the variables in common between this work and
 \citetalias{bernard09a}.}
 \label{fig:deltaper}
\end{figure}

    \subsection[]{Comparison with the ACS data}\label{sec:vimosacs}

 The mean periods of the 506 fundamental mode and 124 first-overtone/double-mode
 RR~Lyrae stars are $\langle P_{ab} \rangle$=0.613$\pm$0.002 d, and $\langle P_{cd} 
 \rangle$=0.381$\pm$0.003 d, respectively. The RR$cd$ represents a fraction of
 $f_{cd} = N_{cd} / (N_{ab}+N_{cd})=0.20$ of the total number of RR~Lyrae stars
 found in this work. These values agree with those found from the ACS 
 survey ($\langle P_{ab} \rangle$=0.614$\pm$0.003 d, $\langle P_{cd}
 \rangle$=0.391$\pm$0.008 d, $f_{cd} = 0.15$), despite
 the differences of location in the galaxy and areas covered. 

 We matched the two catalogues in order to check the consistency of the
 periods found from each data set. Figure~\ref{fig:deltaper} displays the
 difference in period for the 163 RR~Lyrae stars in common, and shows that the
 agreement is excellent. The main outlier at P$\sim$0.64 d is actually a blend of
 two RR$ab$ stars, that are well resolved in the ACS survey, which prevented us
 from recovering their periods correctly. The non-negligible dispersion 
 ($\sigma\sim0.007$ d)
 in the lower panel reflects the fact that the accuracy of the period
 determination in the present survey is limited to $\sim$0.01 d for many of the
 variables.

 In order to refine the periods and mean magnitudes of the RR~Lyrae stars in
 common, we tried to combine the light curves from the two surveys.
 Unfortunately, crowding in the VIMOS data often produced an offset of the
 data points toward brighter magnitudes, as well as reduced the apparent
 amplitude of the pulsations, leading to noisy light curves. In addition, the
 brighter Anomalous Cepheids, which are less affected by these limitations,
 revealed that the temporal sampling is another limiting factor. The short
 length of each observing run compared to the time difference (roughly two years)
 between the two
 runs, combined with the sub-optimal SNR of the VIMOS observations even for the
 bright variables, lead to a comb-like periodogram with dozens of tightly-packed
 peaks with the same significance. Each of these peaks produced nearly identical
 light curves, thus preventing us from obtaining more accurate periods.

\begin{figure}
 \includegraphics[width=8cm]{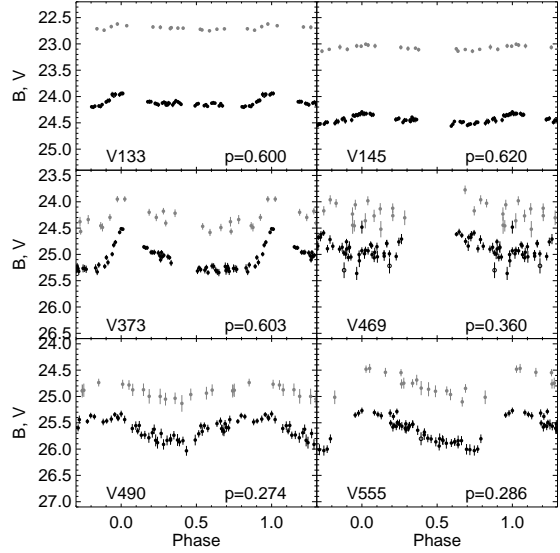}
 \caption{Same as Figure~\ref{fig:rrl_lc}, showing the light curves of a sample
 of RR~Lyrae stars with unusual location on the CMD or in the P-A$_{V}$ diagram.
 See text for details.}
 \label{fig:weird}
\end{figure}

   \subsection[]{Spatial Distribution}\label{sec:distrib}

Figure~\ref{fig:xy} shows the distribution of stars in the field. Open and filled
circles mark the position of the RR~Lyrae stars, while filled triangles show the
Anomalous Cepheids. Noticeably, RR~Lyrae stars are well distributed everywhere in the 
field, clearly tracing the extension of Cetus to about 3.4~kpc (15$\arcmin$) from its 
center, at least along the direction of the major axis. However, despite the number
of variable stars in the outer chips being relatively small, their spatial
distribution appears to be quite asymmetric. In particular, the number
of variables in chip 1, located east of the center, is twice as large as the
number of variables in chip 3, north of the center. Interestingly, the comparison 
with the stellar distribution of \citet{mcconnachie06} presented in Figure~\ref{fig:xy},
shows that the RR~Lyrae 
stars match well the distribution described by the isopleths, which are based on 
the star counts of bright RGB stars. In particular, the outermost isopleth includes
most of the variables detected in chips 1, 2, and 3, despite its truncation in the east
direction.


    \subsection[]{Note on individual stars}\label{sec:note}

In this section we report specific comments on individual variable stars of particular
relevance. In particular, Figure~\ref{fig:weird} shows some of the variable stars 
that have peculiar locations on the CMD or on the P-A$_{V}$ plane. From top to 
bottom, each pair of panels show the light curve of the following stars:

\begin{itemize}
\item {\itshape V133, V145}: These variables belong to the group of RR~Lyrae stars
significantly shifted toward brighter magnitudes and redder colors.
We explain this peculiar location with the
blending of sources, responsible for increasing the brightness and reducing the amplitude
of variable stars. We visually inspected all the suspected blends in common with the 
ACS photometry, verifying that they are contaminated in the lower resolution VIMOS 
images. Note that the vast majority of blended sources are located in chip~4, where 
crowding is the highest. In addition, the fraction of blended to total number of RR~Lyrae 
stars is the same within the area covered by the ACS and over the whole chip~4.
\item {\itshape V373, V469} are the two variables significantly bluer and brighter
than the bulk of RR~Lyrae stars, located at ($B-V,V$)=(0.122,24.307) and (0.146,24.605),
respectively. While the first presents a light curve typical of a RR$ab$ type, the
light curve of the second looks very noisy. Both are blended on the images, and the
morphology of the light curve suggests that they are true RR~Lyrae stars and not faint
Anomalous Cepheids.
\item {\itshape V490, V555} are two examples of RR~Lyrae stars with very short
periods (P=0.274 and 0.286 d). The nature of these objects is not clear (RR$c$ or
RR$e$), but the quality of the light curves clearly supports that the variability
is real. Visual inspection of the images shows that these stars are isolated. The
short period estimated might be an alias of the true period, although it was not
possible to derive good light curves with longer periods for any star belonging to 
this group. Finally, one of these very short-period variable was observed in both 
the ACS survey and this work (V111), and the only period that could phase both datasets
properly is $\sim$0.29 day.
\end{itemize}

\begin{figure}
 \includegraphics[width=8cm]{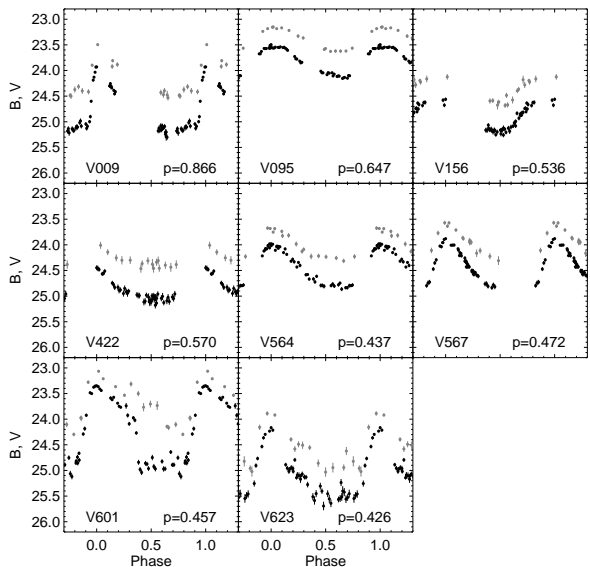}
 \caption{Light curves of the Cepheids in the $B$ (black) and $V$
 (gray) bands, phased with the period in days given in the lower right corner of
 each panel. The $B$-band points were shifted downward by 0.3 mag for clarity.}
 \label{fig:cep_lc}
\end{figure}

\section[]{Anomalous Cepheids}\label{sec:cep}

 Among the variable stars discovered in this survey, eight were identified as
 Anomalous Cepheids (see Table \ref{tab:tab3}). Their light curves are shown 
 in Figure~\ref{fig:cep_lc}. The three Anomalous Cepheids already known from 
 \citetalias{bernard09a}  (V009, V095, V156) were recovered, with periods in 
 good agreement with the  previously determined values. Despite its unusual 
 light curve, V601 is located at the expected position for an Anomalous Cepheid 
 in both the CMD and the period-luminosity (PL) diagram, so in the following we 
 assume it is a bona fide Anomalous Cepheid. Also note that, given the short 
 period measured ($<$1~d), it seems unlikely that these variables are BL Herculis, 
 that is low-mass, old stars evolving from the blue side of the HB.

 To assign the Cepheids to one of the pulsation types (fundamental, F, or first-overtone, 
 FO), one can use either the morphology of the light curve or the location on a PL 
 diagram. The light curves of fundamental mode Anomalous Cepheids are typically 
 very asymmetric, with a sharp rising phase and a slower decline, while they are 
 less asymmetric and closer to a sinusoid for first-overtones. The former are also 
 fainter than the latter at a given period; each type therefore follows a different 
 PL relation. Here the gaps in the light curves partially affected both methods, 
 either by hiding the details of the peak or limiting the accuracy of the calculated
 mean magnitudes.
 Combining the clues from both methods, we tentatively classify V9, V422 and V623
 as fundamental mode Anomalous Cepheids, and the remaining ones as
 overtone pulsators. We note that, with this classification, the three fundamental
 mode Anomalous Cepheids are also the reddest ones, thus in agreement with the fact 
 that they are expected to populate the region of the instability strip close to its
 red edge.

\begin{figure}
 \includegraphics[width=8cm]{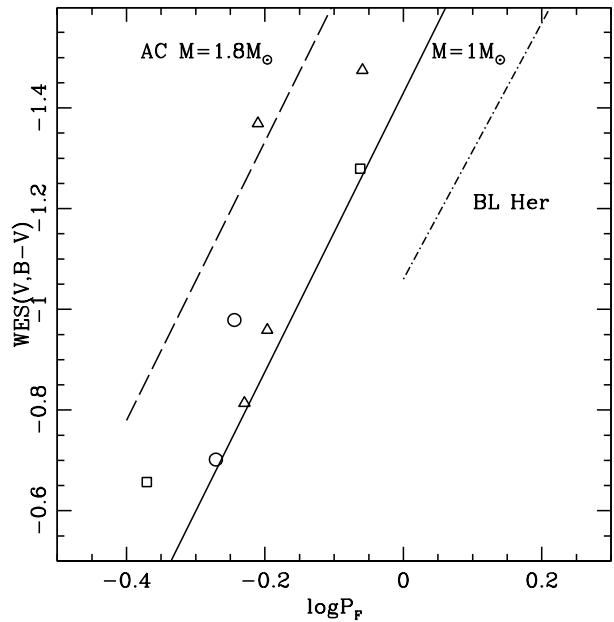}
 \caption{Period-Wesenheit diagram for the Anomalous Cepheids in Cetus. The solid
 and dashed lines show the theoretical locus for stars with masses of 1 and
 1.8~M$_{\sun}$, respectively, while the dotted line shows the expected location of
 BL Her stars. The symbols for the 8 Anomalous Cepheids refer to the classification
 performed with the theoretical relation from \citet{fiorentino06}: the squares are the
 F~mode stars, the triangles the 4 FO, and the circles indicate the two uncertain
 cases.}
 \label{fig:ancep}
\end{figure}

There is an independent way to simultaneously estimate the pulsation mode
and the mass of an Anomalous Cepheid, based on their pulsation properties 
\citep{fiorentino06}, coupled with the period-magnitude-color
({\itshape pmc}) and the period-magnitude-amplitude ({\itshape pma}) relations 
derived in \citet{marconi04}. We recall that, while there are {\itshape pmc} relations
available for both the fundamental ({\itshape pmc\_F}) and first-overtone pulsators
({\itshape pmc\_FO}), the {\itshape pma} relation is valid for fundamental
pulsators only. Applying the three relations to all the stars, we classified as F 
only those for which the {\itshape pma} and the {\itshape pmc\_F} relations provide
consistent mass estimates, within the error bars. The other objects have been classified
as first overtones. With this approach, we could classify 4 FO and 2 F Anomalous
Cepheids, while 2
remain unclear. Remarkably, the classification based on the morphological
and analytical approach coincide for the six stars that have been classified by
both methods.

The estimated masses range from 1.0 to $\sim$1.8 $M_{\odot}$ (see Table \ref{tab:tab3}),
with mean value 1.3$M_{\odot}$. The uncertainty in the mass estimate is of the order 
of 0.1-0.3 $M_{\odot}$. Figure \ref{fig:ancep} presents the reddening$-$free Wesenheit 
($V,B-V$)\footnotemark[14]\footnotetext[14]{ The Wesenheit index is defined 
as $W = V - 3.1\times(B-V)$, where the coefficient comes from the standard
interstellar extinction law \citep{cardelli89}.} 
magnitude, shifted according to the assumed distance, as a function
of the logarithm of the period. The squares mark the stars pulsating in the
fundamental mode, the triangles show the first overtones and the open circles are
the unclear cases. Note that the period of the first overtones have been fundamentalized
adding -0.128 d. The solid and dashed lines present the theoretical relations
for the 1.0 and 1.8 $M_{\odot}$, which appear in excellent agreement with the
masses derived for individual stars. The dash-dotted line shows the analogous
relation for BL Herculis variable stars from \citet{dicriscienzo07}, and supports 
our conclusion that our variable stars do not belong to this group.

At present, two different kinds of progenitors are generally invoked to produce Anomalous
Cepheids: young, relatively massive stars ($<$ 2 Gyr), or BSS originating from primordial
binaries. Excluding the first hypothesis on the basis of the SFH \citepalias{monelli10b}
implies that we should expect all the observed Anomalous Cepheids to be the progeny of BSSs.
The masses derived for our sample of Anomalous Cepheids (from 1.0 to $\sim$1.8 $M_{\odot}$ 
with mean value 1.3$M_{\odot}$),  support this occurrence. In fact, they are 
in excellent agreement with the mass derived for the BSSs \citep{monelli12}, which
is of the order of 1.0-1.3 $M_{\odot}$, and in any case no larger than twice the
mass of the stars evolving at the TO ($\approx$1.6$M_{\odot}$). It is also worth
noting that Anomalous Cepheids are expected to form only in metal-poor populations
\citep[$Z\la$0.0008;][]{fiorentino06}. The low mean metallicity derived from the 
SFH ($\sim$0.0005) is in agreement with this expectation.

\begin{figure}
 \includegraphics[width=8cm]{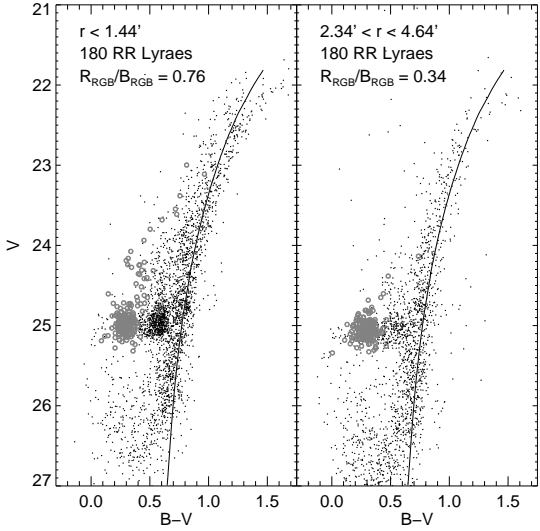}
 \caption{CMDs of the inner and outer regions of Cetus, selected to contain the
 same number of RR~Lyrae stars. A 12~Gyr old isochrone with Z=0.0003 from the
 BaSTI library is over-plotted to guide the eye. RR~Lyrae stars are shown as
 larger gray symbols.}
 \label{fig:cmd_inout}
\end{figure}


 \section[]{Radial gradients}\label{sec:radial}

 Given the large area of the galaxy covered by the present observations, we can now
 search for gradients in the properties of the stellar populations of Cetus. In particular,
 we will focus on the HB, RGB, and RR~Lyrae stars, in order to check if possible spatial
 changes with the stellar population properties are reflected in the properties
 of RR Lyrae stars, as observed in Tucana \citep{bernard08}.

 In \citetalias{bernard09a}, we noticed a small but significant gradient of the
 HB morphology parameter in Cetus, in the sense that the color spread of the HB 
 is larger in the
 center, as in Tucana. However, the limited field-of-view of the ACS as well as the
 relatively low number of variable stars prevented us from asserting the
 veracity of the gradient or detecting a possible change in the mean
 periods of the RR~Lyrae stars.

In the following, we will use different stellar tracers to identify radial 
gradients in Cetus and to answer the question: is the old population in Cetus
similar to that of Tucana, with two clear sub-populations with different properties?

	\subsection[]{RGB color distribution}

Taking advantage of the large field-of-view of the VIMOS camera, we first investigate
if the color distribution of the RGB changes with radius. Figure~\ref{fig:cmd_inout} compares
the CMD of the inner ($r < 1.44\arcmin$) and outer ($2.34\arcmin < r < 4.64\arcmin$)
regions, selected to contain the same number of RR~Lyrae stars. Using the 12
Gyr isochrone over-plotted on both panels as reference, a clear difference appears
between the two RGBs. The one in the central region presents a larger number of stars
on the red side, and a larger color dispersion, especially at the bright end. The
blue side looks well populated in both diagrams. To support this occurrence,
we selected two regions on the blue and red side of the aforementioned isochrone, and
estimated the number of star brighter than $V$=24.5~mag in both CMDs, carefully avoiding
the pollution from HB and AGB stars. The ratio of
the red to blue stars changes by 0.76 to 0.34 from the inner to the outer region.
This supports our earlier conclusion that the more metal-poor population is
equally present all over the galaxy body, while the more metal-rich one, responsible
for the reddest RGB stars, is preferentially concentrated in the innermost regions.
This is in agreement with what is commonly observed in many dwarf galaxies in the
LG \citep[e.g.,][]{bellazzini01,harbeck01,tolstoy04,battaglia06}.

        \subsection[]{HB morphology}\label{sec:hbr}

\begin{figure*}
 \includegraphics[width=15cm,height=10cm]{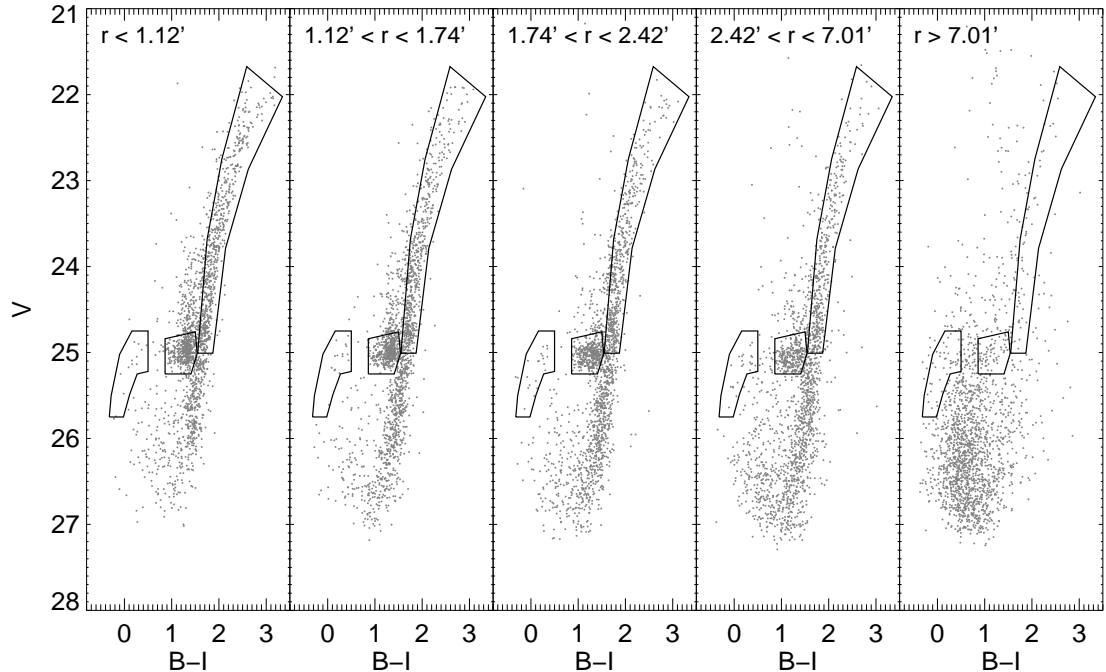}
 \caption{($V$, $B-I$) CMDs for five concentric regions containing the same total
 number of stars. The boxes used to count the number of blue HB, red HB,
 and RGB stars are over-plotted.}
 \label{fig:hbr_boxes}
\end{figure*}

To quantitatively study the shape of the HB we used the HB ratio ($HBR$\footnotemark[15]),
defined counting the blue, variable, and
red\footnotetext[15]{$HBR$ = (B$-$R)/(B+V+R), where B, V, and R are the numbers of
stars to the blue, within, and to the red of the instability strip \citep{lee90}.}
stars in the boxes presented in Figure~\ref{fig:hbr_boxes}. The selection was performed
in the ($V$, $B-I$) plane for two reasons. First, the larger color baseline
helps to separate the features in the CMD, specially limiting the overlap
between the red HB and the RGB. Second, the $B$ and $I$ colors are similar to the
$F475W$ and $F814W$ of the ACS data. This allowed us to check our selection in the
ACS photometry, verifying that no relevant stars were left outside the box, especially 
blue stars, while at 
the same time keeping the number of polluting objects included in the selection low.

In total, we identified 111 and 1527 blue and red HB stars, respectively. Given 
the 630 RR~Lyrae stars identified, this gives a global $HBR$=$-$0.62$\pm$0.01.
If we limit this estimate to the area in common with the ACS only, we find 
$HBR$ = $-$0.67$\pm$0.03, in good agreement with that found from the ACS photometry 
\citepalias[i.e., $-$0.74;][]{bernard09a}. Taking into account the higher spatial 
resolution of this camera, and the better star-galaxy separation, this comparison 
reinforces the conclusion that our cleaning of the VIMOS CMD was effective in this 
magnitude range.

\begin{figure}
 \includegraphics[width=8cm]{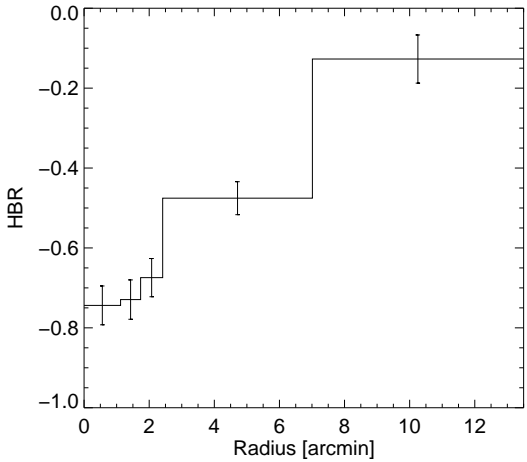}
 \caption{Variation of the $HBR$ as a function of galactocentric radius.
}
 \label{fig:hbr_radius}
\end{figure}

However, more interesting results are found when studying the HBR variation as a 
function of radius. To do this, we divided the clean list of stars into five regions,
containing the same total number of stars. Figure~\ref{fig:hbr_boxes} presents the
CMDs of the five regions. Note that the constraint on the number of stars implies
that the area of the ellipses is different, the outermost one being $\sim$37 times
larger than the innermost one. This also explains the increase in the number of
blue objects fainter than the HB for increasing galactocentric radius. For this
reason, the value of the HBR may not be reliable in the last radial bin.
The $HBR$ variation is presented in Figure~\ref{fig:hbr_radius}.
The plot shows a clear trend: the larger the distance from the center of the galaxy,
the higher the value of the $HBR$, which changes from $-$0.75 to $-$0.1. 
This means that the HB gets bluer with increasing distance.  This is due to the fact that 
the slightly younger stars populating the red HB are more centrally concentrated. 
This is in excellent agreement with the variation of the RGB thickness with radius,
and similar to what was found in Tucana \citep{bernard08}.

	\subsection[]{Properties of the RR~Lyrae stars}\label{sec:var_grad}

\begin{figure}
 \includegraphics[width=9cm]{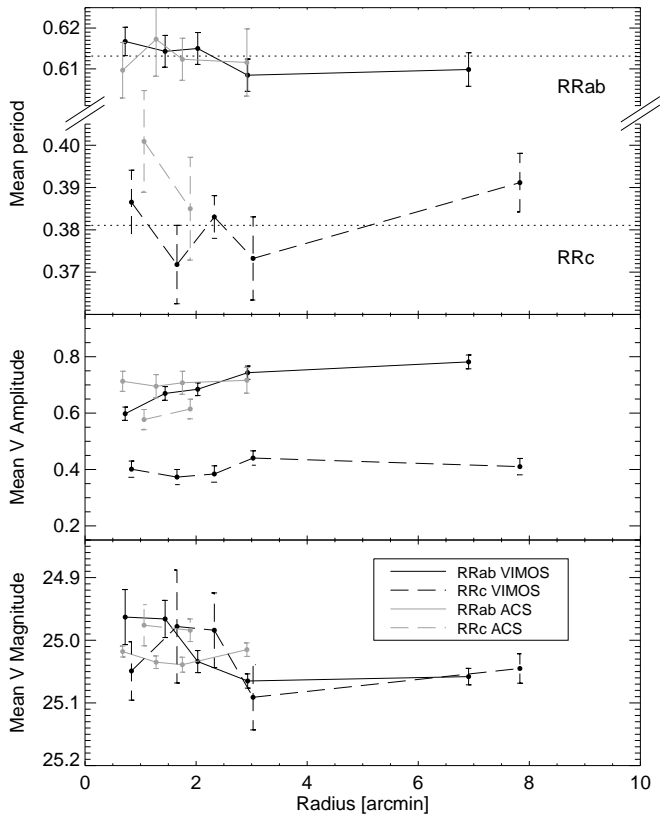}
 \caption{ Radial variations of different quantities for the RR$ab$ (solid lines)
 and RR$c$ stars (dashed line). The black and gray lines show the stars from
 the VIMOS and ACS surveys, respectively. The bins have been chosen such as to contain 
 the same number of stars. The three panels show, from top to bottom, the variation
 of the mean period, the mean V amplitude and the mean V magnitude. See text for details.}
 \label{fig:pergrad}
\end{figure}

In this section we investigate if the pulsational properties of the RR~Lyrae stars 
of Cetus present any variation with galactocentric radius. Figure~\ref{fig:pergrad} 
shows for both the RR$ab$ (solid black line) and RR$c$ (dashed black line) variables,
the mean period, mean $V$ amplitude, and the mean $V$ magnitude as a function of radius. 
For each profile, the samples have been selected to contain the same number of variables 
in each bin. The upper panel shows no obvious radial trend of the mean period. As a 
comparison, the gray lines show analogous plots based on the ACS measurements, disclosing 
the same behavior. The mean $V$ amplitude and magnitude show a mild trend, in the sense of 
decreasing amplitudes and increasing (dimming) magnitudes with increasing radius. However,
this can be explained with the increasing crowding toward the innermost regions, which 
artificially pushes stars to brighter magnitude (see also Figure~\ref{fig:cmd_vars}) and
also reduces the measured amplitude. This interpretation is supported by the comparison
with the results from the ACS data (gray lines), which do not present a similar gradient, 
but rather are consistent with a flat distribution within the errors. 

 As a further check on the possible presence of period gradients, we verified the
 radial distribution of the RR$ab$ stars located on either side of the solid
 black line fitting the P-A$_{V}$ relationship shown in the top panel of Figure~\ref{fig:pa_hist}.
 The reason for this test is that the location of the RR$ab$ stars in the P-A$_V$
 diagram is a function of metallicity, in the sense that more metal-rich stars
 tend to have shorter periods and lower amplitudes \citep{dicriscienzo04,sandage04,kunder11}.
 The resulting profiles are virtually  indistinguishable, 
 implying that the mean period of the RR$ab$ stars is homogeneous over the whole galaxy.

We conclude that all of this evidence indicates that the pulsational properties of the 
RR~Lyrae stars are constant, within errors, over the whole body of Cetus, 
despite the significant change of the HB morphology.


\section[]{Discussion}\label{sec:discussion}

\subsection[]{Comparing Cetus and Tucana}\label{sec:comparing}

Cetus and Tucana are the two most isolated dSph galaxies in the LG. 
Both consist mostly of stars older than 9 Gyr and more metal-poor than
Z=0.001. Both galaxies present evidence of stellar population gradients, in
the sense that the RGB color spread and the HB morphology change as a function of
galactocentric radius. From the ACS data, for Tucana it was clear  
that the red component of both features vanishes faster, with increasing radius, 
than the blue one. However, for Cetus, this finding was
not as firmly established. \citetalias{bernard09a} had shown hints of
gradients in both the HB and RGB morphologies of Cetus, but the limited area covered
by the ACS, extending from 0.5 to $\sim$3 core radii, prevented 
solid conclusions.  Due to the large field-of-view of the VIMOS camera, 
the present photometry supports these findings.  In particular, we
can confirm the presence of gradients in the HB and RGB of Cetus similar to those
of Tucana: the red component of both features is more centrally concentrated
than the blue one. Nevertheless, the HB is populated at both sides of the instability
strip at any radius. This suggests a more prolonged or more efficient star formation in the
innermost regions, implying that the star forming region shrinks toward
the center with time. Detailed SFHs calculated at different galactocentric
radii (\citealt{hidalgo11b}, Hidalgo et al. in prep) supports this scenario.
In fact, it shows that the {\it peak} of star formation occurred at the same
age all over the ACS field, but the star formation was more extended, and
produced the most metal-rich stars closer to the center.

Despite the overall similarity, striking differences also appear in the
Cetus-Tucana comparison. The most obvious are the different HB
morphologies and the properties of the RR Lyrae stars. Tucana has an extended
and well populated HB from the blue to the red, with the most peculiar
feature of a clear split in magnitude at intermediate colors. This is
reflected in the presence of two populations of RR Lyrae stars with
different properties and radial distributions, and two sequences of redder
stars. This also has a clear counterpart in two RGB bumps \citep{monelli10a}.
In contrast, the HB of Cetus is more evenly populated in magnitude; it is overall
redder than that of Tucana and has a sparsely populated blue side. Only
one RGB bump was detected in Cetus. In this paper, we extended the sample of RR
Lyrae stars by a factor $\sim$3.7 with respect to that presented in
\citetalias{bernard09a}. We substantially confirm the global
properties derived on the basis of the smaller and central ACS sample,
which pointed to the conclusion that the properties of the Cetus RR Lyrae
were homogeneous over the whole area studied. In particular, we did not
detect any statistically significant variation of the mean period,
amplitude, or mean magnitude of the RR Lyrae stars as a function of the
galactocentric radius. We also verified that the slope of the P-A$_{V}$
relation does not change with radius. Interestingly, this relation shows
two noteworthy properties: {\em i)} the steep slope -- and linearity
-- of the relation, compared to the shallower relation typically found in
Galactic GCs and LG galaxies; and {\em ii)} the very small dispersion
around the mean. While we cannot explain the former, the latter further
supports the conclusion that the properties of RR~Lyrae stars are indeed
homogeneous in the field. {\it All of the evidence suggests that the RR~Lyrae 
stars in Cetus formed from a parent population with homogeneous
properties in terms of a small range of age and metallicity.} 

It is interesting to verify whether the RR Lyrae population of Cetus presents
similarities with that of Tucana. We estimated the mean $V$ 
magnitude and dispersion of the
RR Lyrae stars by fitting the magnitude histogram. In this case, we used
the ACS \citepalias[table 7 and 8,][]{bernard09a} sample because of the bias toward 
brighter magnitudes caused by crowding in the VIMOS data. However, note that
this only affects the spread in the VIMOS measurements, because we found that 
the mean magnitude is consistent with the one from the ACS data within 0.02 mag.
The Gaussian profile fit to each histogram gives mean magnitudes in excellent agreement 
between the two galaxies:$<M{_V}>$ = +0.48 and +0.47 for Cetus and Tucana, respectively, 
with dispersions of $\sigma$=0.07 and 0.09~mag. Interestingly, if we split the 
sample of Tucana RR Lyrae stars into a {\it bright} sample and a {\it faint} sample,
following \citet{bernard08}, we find that Cetus has intermediate properties
between the two. In fact, the derived mean magnitudes for the bright and the
faint groups are $<M{_V}>$ = +0.41$\pm$0.05 ~mag and $<M{_V}>$ =
+0.54$\pm$0.04~mag, respectively.
Moreover, while both galaxies can be classified as Oosterhoff-intermediate
objects on the basis of their global properties, \citet{bernard08} noted
that the two sub-populations of Tucana are closer one to the Oosterhoff I
type, and the other to the Oosterhoff II type. This is not the case for Cetus,
where the pulsational properties of the bright and faint samples 
are indistinguishable. Finally, neither of the two P-A$_{V}$ relations of 
Tucana resemble that of Cetus, which is significantly steeper than both. 
We conclude that the population of Cetus RR Lyrae stars has no direct counterpart in Tucana.

\subsection[]{The SFHs of Cetus and Tucana}\label{sec:sfhs}

We can try to explain this by correlating all the available evidence with
the obtained SFH. We should bear in mind that the main effect of the
observational errors, especially at old ages, is mostly to broaden the
estimated duration of the events of star formation. Our technique to
estimate this effect \citep[see][ in particular Fig. 10]{monelli10c} 
disclosed that the main event of star formation lasted between 1 and 3 Gyr
for both galaxies. Moreover, the star formation rate was significantly
stronger in Tucana than in Cetus 13 Gyr ago, while it was the opposite 1
Gyr later. At some point in between, the two galaxies had comparable 
activity. This is reflected in the present-day properties that we observe.
It must be stressed that, in general, the analysis of the HB morphology is a
very complicated matter \citep[see, e.g.,][and references therein]{gratton10}
that depends not only on the evolution of
stars (which in turn depends on age, metallicity, chemical peculiarities in
the He and CNO abundances), but also on uncertainties introduced by
other mechanisms like the mass loss during the RGB phase (as widely
discussed for globular clusters). However, there is a fundamental
difference between globular clusters and dwarf galaxies like Cetus and
Tucana. While the formation of a cluster is a very rapid event, resulting
in a very limited age distribution, in the case of a dSph, the star
formation occurred over a much longer period of time. Moreover, in general,
dSphs experienced important early chemical evolution. In particular, both
Cetus and Tucana present a well defined age-metallicity relation
\citep{monelli10c}. These two observations together imply that the
stars populating the HB at the present time must be the evolution of
progenitors characterized by considerable spreads in age and 
chemical composition. It is, therefore, reasonable to expect that, in dSph
galaxies, the two parameters age and metallicity, dominate over other
possible parameters influencing HB morphology.  The above provides
reasonable explanation for all of the observations of Cetus and Tucana. 
The following discussion enters a more speculative realm. 

Starting from this scenario, and assuming standard mass loss on the RGB
over the relevant time period, we would expect a general ranking along 
the HB. If the HB we observe today is the
result of the superposition of simple stellar populations with different
ages and metallicity, it is reasonable to suppose that the blue HB stars
would proceed from more metal-poor, older and less massive progenitors, 
while the reddest ones would be younger, originally more massive and, due
to the chemical evolution of the host galaxy, more metal-rich. Therefore,
the very strong initial burst in Tucana might have produced an important
population of metal-poor stars, which are expected to populate the
extended blue HB (and correspondingly the blue RGB), the population of both 
bright RR Lyrae and bright red HB stars and the brighter RGB bump. 
This first event, in turn, caused an
important self-enrichment on very short time scales, fast enough to have a
detectable and discrete change in the mean metallicity, which appears in
the faint population of RR Lyrae stars, the faint red HB (and
correspondingly the red RGB), and the second RGB bump.
Following the same reasoning, the lower initial activity in Cetus would
explain the few blue, old and metal-poor HB stars, while the subsequent
stronger star formation might be responsible for the overall redder HB
morphology. As in the case of Tucana, Cetus shows an age-metallicity
relation, but the metallicity evolution must have been such that the
metallicity increment during the time in which the progenitors of the
present day RR Lyrae stars were formed was not enough to produce RR Lyrae 
with a range of properties as wide as those of Tucana.

All this evidence suggests that the RR Lyrae stars in these two dSph are
not representative of, generically, the {\itshape whole} ``old'' population,
but are rather a ``picture'' of a well defined sub-population (one in the
case of Cetus, two in the case of Tucana) characterized by a relatively
small age and metallicity dispersion, in the older portion of the population 
of each galaxy. This would also be in agreement with the fact that the two 
groups of RR Lyrae stars in Tucana have different radial distributions: 
the faint, more metal-rich stars of the second generation are more 
concentrated to the center.

We stress that this scenario is certainly simplified and, at some level,
conjectural. DSph galaxies are more complex systems than globular clusters,
and a full modeling of the HB, taking into account all the relevant parameters
affecting its morphology, would be required for further insight.
However, note that a complete age-metallicity degeneracy cannot be at
work, otherwise we would not detect the discrete nature of the Tucana HB,
but rather a continuous distribution of stars.

What physical processes led to the differences in evolution between the two
galaxies is not immediately obvious.  Two possible mechanisms could be (i) 
differences in early gas infall, or, in the case of Tucana, (ii) the two populations
may be due to a very early merger.  Perhaps the most important point is that
the early evolution histories of these two systems are not identical.  Even
systems this small show a variety in their early evolution.


\section{Conclusions}\label{sec:conclusions}

We analyzed time-series $B$,$V$ VLT/VIMOS data of the isolated dSph galaxy Cetus.
The main results of this work are:

\begin{itemize}

\item we detected 638 variable stars, all bona fide members of Cetus: 630 RR~Lyrae 
stars (470 new discoveries) and 8 Anomalous Cepheids (5 new). We provide a complete 
catalogue of periods, amplitudes, and mean magnitudes;
\item we confirm the results presented in \citetalias{bernard09a} concerning the mean 
pulsational properties of RR~Lyrae stars. In particular, Cetus is, similar to 
other satellites dSph galaxies, of the Oosterhoff intermediate type. The P-A$_{V}$ 
relation is linear, and presents a particularly steep slope and limited dispersion, 
similar to that of Carina;
\item The wide field of view of the VIMOS camera allowed us to detect a clear gradient 
in the HB and RGB morphologies. In particular, both features get bluer, on average, 
for increasing distance from the center of Cetus, due to the steeper radial profile of the 
red component;
\item On the contrary, the pulsational properties of RR~Lyrae stars did not reveal any 
obvious gradient. The mean $V$ magnitude, the mean $V$ amplitude, and the mean period are 
consistent with a flat distribution as a function of radius;
\item The homogeneity of the RR~Lyrae population of Cetus is 
different from that of Tucana. We interpret this in concordance with the SFHs derived for 
the two galaxies (Paper II, \citealt{monelli10c}). Different observables strongly suggest
the presence of two distinct old populations in Tucana, with the second slightly younger
and slightly enriched from the ejecta of the first one, while Cetus had a smoother
evolution reflected in more homogeneous present-day properties. Nonetheless,
it is interesting to stress that the pulsation properties of RR~Lyrae stars in Cetus
do not resemble either population of Tucana, being a purely intermediate Oosterhoff type.
All of the evidence also suggests that, despite the similar ranges of age and metallicity 
of the stellar populations in Cetus and Tucana, the details of the differences in early
SFH clearly show up in the morphology of the HB and the properties of RR Lyrae stars.
\end{itemize}

\section*{acknowledgments}

\small
We thank Michele Cignoni for performing simulations of the Galactic field using
the Pisa Galactic Model code, and the anonymous referee for useful comments.
Support for this work was provided by the IAC (grant 310394), the Education and
Science Ministry of Spain (grants AYA2007-3E3506, and AYA2010-16717), and a
rolling grant from the UK Science and Technology Facilities Council. S.C. warmly thanks
INAF for the financial support through the PRIN INAF 2009 (PI: R. Gratton).
\normalsize

\bibliography{monelli_bibtex.bib}

\appendix
\section{Finding Charts}

\begin{figure*}
 \includegraphics[width=17cm]{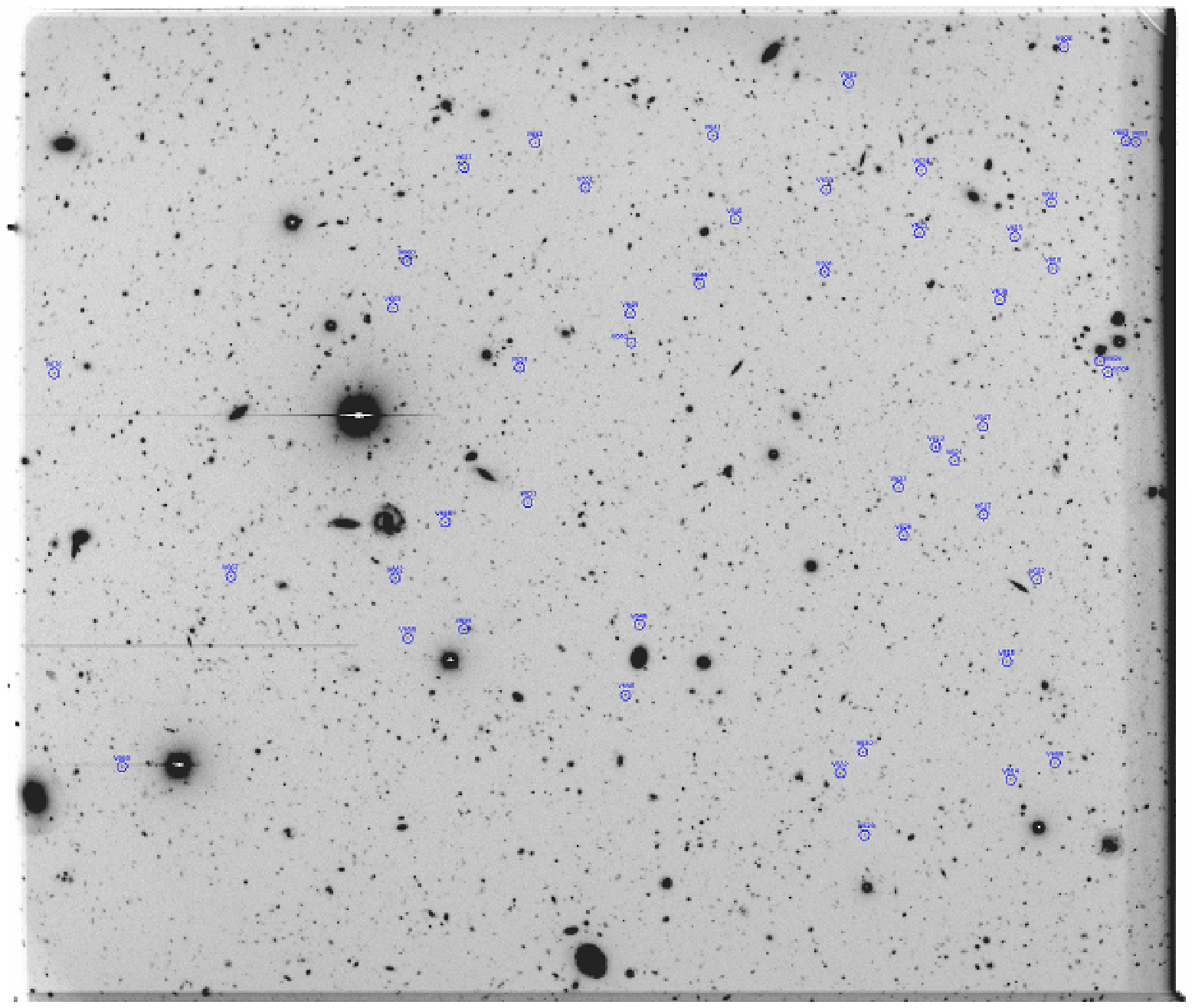}
 \caption{Finding chart for variables stars in Chip~1. The full set of
 high-resolution finding charts is available in the electronic edition.
 \label{lastpage}}
 \label{fig:fc1}
\end{figure*}

\end{document}